\documentclass[reqno]{amsart}

\usepackage[margin=3cm]{geometry} %set the margins

\usepackage{bm}
\usepackage{hyperref} %[hidelinks] hides the fluorescent boxes and makes all colors black
\hypersetup{
    colorlinks = true,
    linkcolor={red!50!black},
    citecolor={blue!50!black},
    urlcolor={blue!50!black}
}
\usepackage[utf8]{inputenc}
\usepackage{cite}

\usepackage{amsmath,accents}
\usepackage{siunitx}
\usepackage{mathtools}
\usepackage{amssymb}
\usepackage[normalem]{ulem}
\usepackage{enumitem}
\usepackage[capitalise]{cleveref}

\usepackage{tcolorbox} %allows one to put colored boxes around text and equations (cannot be used with xcolor package}

\usepackage{booktabs,siunitx,array,threeparttable}
\def\be{\begin{equation}}
\def\ee{\end{equation}}
\def\e#1{\label{#1}\end{equation}}
\def\bea{\begin{eqnarray}}
\def\eea{\end{eqnarray}}
\def\ea#1{\label{#1}\end{eqnarray}}
\def\bse{\begin{subequations}}
\def\ese{\end{subequations}}

\def\bem#1{\begin{mathletters}\label{#1}}
\def\eml{\end{mathletters}}

\def\est#1{\hat{#1}}

\newcommand{\Op}[1]{\bm{#1}} %Definition of operator notation

\newcommand{\ket}[1]{\left| #1 \right>}
\newcommand{\bra}[1]{\left< #1 \right|}

\newcommand{\iu}{{i\mkern1mu}}

\newcommand{\myvec}[1]{\accentset{\rightharpoonup}{#1}}
\def\4#1{{\boldsymbol{#1}}}
\def\8#1{\mathbb{E}\left[#1\right]}
\def\func#1#2{\mathrm{#1}\left[#2\right]}
\def\funcwrt#1#2#3{\mathrm{#1}_{\mathrm{#3}}\left[#2\right]}

\DeclareMathSymbol{\mlq}{\mathord}{operators}{``}
\DeclareMathSymbol{\mrq}{\mathord}{operators}{`'}

\usepackage{titlesec} %titlesec package
\titleformat{\subsection}{\centering\normalfont\bf}{\thesubsection.\quad}{0em}{}{}
\renewcommand{\thesubsection}{\arabic{subsection}}

\newtheorem*{contradiction}{The contradiction}

\theoremstyle{definition}
\newtheorem{definition}{Definition}
\newtheorem{remark}{Remark}

\makeatletter
\newcommand{\myitem}[1]{%
\item[#1]\protected@edef\@currentlabel{#1}%
}
\makeatother

\begin{document}

\title{Quantum squeezing cannot beat the standard quantum limit}
%\title{Squeezed states cannot beat the standard quantum limit or the single particle limit}

\maketitle

\begin{centering}
\author{Liam P. McGuinness}\\
\address{Laser Physics Centre, Research School of Physics, Australian National University, Acton,}\\
\address{Australian Capital Territory 2601, Australia}\footnote{Currently at Institute for Quantum Optics, Ulm University, Albert-Einstein-Allee 11, D-89081 Ulm, Germany.}\\
\email{Email: \href{mailto:liam@grtoet.com}{liam@grtoet.com}}\\
\end{centering}

%\author{Liam P. McGuinness}
%\email[Email: ]{liam@grtoet.com}
%\affiliation{Laser Physics Centre, Research School of Physics, Australian National University, Acton, Australian Capital Territory 2601, Australia}

\begin{abstract}
Quantum entanglement between particles is expected to allow one to perform tasks that would otherwise be impossible \cite{Bell1964, Feynman1982, Deutsch1985, Shor1994, Giovannetti2001}. In quantum sensing and metrology, entanglement is often claimed to enable a measurement precision that cannot be attained with the same number of particles and time, forgoing entanglement \cite{Helstrom1969, Holevo1973a, Braunstein1996, Giovannetti2004, Giovannetti2006, DemkowiczDobrzanski2012, Zwierz2012, Pezze2018}. Two distinct approaches exist: creation of entangled states that either \emph{i)} respond quicker to the signal, or \emph{ii)} are associated with lower noise and uncertainty. The second class of states are generally called squeezed states. Here we show that if our definition of success is a precision that is impossible to achieve using the same resources but without entanglement then squeezed states cannot succeed. In doing so we show that a single non-separable squeezed state provides fundamentally no better precision, per unit time, than a single particle.\\
\end{abstract}

%\emph{... as men filled not with wisdom but with the conceit of wisdom they will be a burden to their fellows}
%-- Plato, \emph{Phaedrus} (Oxford World’s Classics, 2009)

\noindent \emph{You have nothing to do but mention the quantum theory, and people will take your voice for the voice of science, and believe anything.}
--- George Bernard Shaw, \emph{Geneva} (1938)\footnote{Hat-tip Daniel Simon. ``On the Power of Quantum Computation''. \emph{SIAM Journal on Computing}. \textbf{26}, 1474--1483. (1997).}

\section*{Few sentence summary}
A quantum state is a unit length ray in a complex Hilbert space. The sensitivity of a quantum state is determined \emph{solely} by how much the state changes in response to a change in a parameter $\theta$. This bounds the information contained in any measurement and ultimately the precision of any estimate of $\theta$. In contradiction, squeezed states are generally claimed to improve the quantum Fisher information without increasing the state response to $\theta$, they do so by reducing or squeezing noise.

%The quantum Fisher information (on a parameter $\theta$) of a quantum state is determined by how much the state responds to a change in $\theta$. This in turn bounds the classical Fisher information and ultimately the precision of any estimate of $\theta$. In contradiction, squeezed states are claimed to improve the quantum Fisher information without increasing the state response to $\theta$.

\section*{Summary}
In quantum metrology, a standard formula bounds how much information a single quantum state can provide on an unknown parameter $\theta$ during sensing time $t$. It is:
\[ \func{I}{\theta} \leq \left(t \cdot E_{\delta}\right)^2. \]
In words, for a given time the information on $\theta$, contained in a quantum state, is bound by $E_{\delta}$, the maximum difference in energy eigenvalues between two eigenstates of $\ket{\psi}$.\\

The power of this relation is it's simplicity. When characterising the sensitivity of a given quantum state the only thing that matters is how much the state responds to the signal: $E_{\delta}$ sets an upper bound on this. If the bound can be saturated and any equal superposition of eigenstates differing in their eigenvalues by $E_\delta$ reaches the bound, then there is one way, and \emph{only one way} to increase the information contained in a quantum state.\footnote{The main text focusses on showing that a single spin can saturate the bound and highlights some of the incorrect and misleading analysis that is common in the field.} The equation can be restated in a different form. What properties must a quantum state have to achieve a better precision? It must have a greater change in Hilbert space angle and that is achieved by increasing $E_{\delta}$; the energy gap between the superposition states composing $\ket{\psi}$. When designing better quantum sensors this relation is particularly instructive. A better sensor is one that is a superposition over maximally different energy eigenstates. This quantum state responds more to changes in $\theta$ and is more sensitive than a state with lower energy gap (value of $E_{\delta}$). In quantum metrology, these more sensitive states are created by entangling $N$ spins to produce a state with $N$-fold increased energy gap and are commonly known as NOON, CAT or GHZ states. Such a state is equivalent to a single spin-$1/2$ but with $N$ times greater magnetic moment.

\begin{contradiction}
Squeezed states do not have this basic property.
\end{contradiction}
As commonly described, a single squeezed state does not have an increased energy gap when compared to say a single spin. See the review articles covering squeezing \cite{Ma2011, Gross2012, Ludlow2015, Pezze2018}, Refs.\,\cite{Caves1981, Walls1983, Wodkiewicz1985, Wu1986, Slusher1987, Xiao1987, Polzik1992, Wineland1992, Kitagawa1993, Wineland1994, Sanders1995, Kuzmich1998, Soerensen1998, Brif1999, Kuzmich2000, Meyer2001, Orzel2001, Andre2002, Esteve2008, Appel2009, Eberle2010, Gross2010, Koschorreck2010, Koschorreck2010a, Leroux2010, Leroux2010a, LouchetChauvet2010, Riedel2010, SchleierSmith2010, Wasilewski2010, Abadie2011, Luecke2011, Sewell2012, Aasi2013, Taylor2013, Bohnet2014, Muessel2014, Strobel2014, Bohnet2016, Cox2016, Davis2016, Hosten2016a, Hosten2016, Kruse2016, Linnemann2016, Macri2016, Polzik2016, Braverman2019, Tse2019, Bao2020, Malia2020, PedrozoPenafiel2020, Schulte2020, Casacio2021, Gilmore2021, Colombo2022, Greve2022, Malia2022, Bornet2023, Eckner2023, Franke2023} and \href{https://www.nist.gov/quantum-information-science/quantum-sensing-explained/squeezed-light-detecting-gravitational-waves}{recent} \href{https://physics.aps.org/articles/v16/189}{press} \href{https://www.ligo.caltech.edu/news/ligo20231023}{releases} on frequency dependent squeezing in LIGO \cite{Ganapathy2023,Jia2024}. Supposedly, squeezed states provide enhanced precision by reducing some fundamental noise, and \emph{not} by increasing the state response. I agree that this argument sounds reasonable. It is persuasive because it appeals to our intuitive concept of signal-to-noise determining the measurement precision. I refer you to the above formula -- based on standard quantum mechanics this is impossible. An increase in $E_\delta$ is a necessary condition to obtain more information than another state such as a single spin which saturates the bound. Any state with the same $E_\delta$ obeys the same information bound, even allowing that it has ``less noise'' due to squeezing.\\

%\section*{The contradiction}
To restate the above formally, let $\ket{\psi}$ be a single, \hyperlink{non-separable}{non-separable} quantum state that interacts with a time-independent Hamiltonian of the form $\theta\cdot\Op{H}$. Then after interaction time $t$, the \hyperlink{QFI}{quantum Fisher information} on $\theta$ of this state is bounded by
\be \func{QFI}{\ket{\psi},\theta} \leq \left(t \cdot E_{\delta}/\hbar\right)^2 \label{eq:QFI}, \ee
where $E_{\delta}$ is the maximum eigenvalue difference between any two energy eigenstates comprising $\ket{\psi}$, $\hbar$ is Planck's constant\footnote{Often $\hbar$ is set to one.} and it is assumed that initially $\ket{\psi}$ has no dependence on $\theta$. The above analysis applies to almost all demonstrations using squeezed states, including gravitational wave detection (in the limit the detection timescale/bandwidth is short compared to the period of the gravitational wave). \Cref{eq:QFI} is a standard bound in quantum metrology, %\footnote{Equivalent analysis to \cref{eq:QFI} is presented the following references: Ref.\,\cite{Giovannetti2006}-Eqs.\,(4--7); Ref.\,\cite{Zwierz2012}-Eq.\,(8--10); Ref.\,\cite{Giovannetti2011}-Eq.\,(5); Ref.\,\cite{Boixo2007}-Eq.\,(2); Ref.\,\cite{Tsang2012}-Eq.\,(11); Ref.\,\cite{Pang2014}-Eqs.\,(7,\,11); Ref.\,\cite{Jarzyna2015}-Eq.\,(2); Ref.\,\cite{Yuan2015}-(Eq.\,(7) and introduction); Ref.\,\cite{Pang2017}-Eq.\,(7); Ref.\,\cite{Gorecki2020}-Eqs.\,(1--2); Ref.\,\cite{Len2022}-Eq.\,(4).} 
more detail on it's widespread appearance in the literature is provided in Appendix\,\ref{App:refs} and it's extension to arbitrary time-dependent Hamiltonians is given in Appendix\,\ref{App:General}. % and it is built upon the foundational work of Helstrom \cite{Helstrom1976} and Holevo \cite{Holevo1982}.
\\

\Cref{eq:QFI} is important when $\ket{\psi}$ is used as a sensor to estimate the value of $\theta$. Let $x$ be a single measurement datum sampled from probability distribution $\func{Pr}{x;\theta}$. Classical estimation theory tells us the Fisher information on parameter $\theta$ is quantified by the gradient of the (logarithm of the) probability distribution with respect to $\theta$ \cite{Kay1993} (Ch.\,3):\footnote{Here $\8{\cdot}$ denotes expectation and is taken with respect to $\func{Pr}{x;\theta}$.}%, the derivative is calculated at the true value of $\theta$. The equality holds assuming that we use an unbiased estimator, the measurement operator $\Op{X}$ corresponding to the observable $x$ does not depend on $\theta$, and we can interchange the trace and the derivative to write $\frac{\partial}{\partial\theta}\func{Tr}{\rho \Op{X}} = \func{Tr}{\frac{\partial \rho}{\partial \theta} \Op{X}}$.}:
\be \func{I}{\theta} := \8{\left(\frac{\partial \ln\left[\func{Pr}{x;\theta}\right]}{\partial \theta}\right)^2}.\label{eq:CFI} \ee
The relevance of quantum mechanics for this bound is the following. Given a single \hyperlink{non-separable}{non-separable} quantum state $\ket{\psi}$ with \hyperlink{QFI}{quantum Fisher information} $a$, then $a$ also bounds the classical \hyperlink{Fisher information}{Fisher information} of any measurement datum $x$ obtained from $\ket{\psi}$. I.e. given a quantum state, the Fisher information of any realizable measurement on this state is bound by the quantum Fisher information %no measurement allowable by quantum mechanics can produce a datum $x$ with classical Fisher information greater than $a$
\[ \func{I}{\theta} \leq \func{QFI}{\ket{\psi},\theta} \leq \left(t\cdot E_\delta/\hbar\right)^2.\]
Furthermore, in statistical parameter estimation the Fisher information bounds the mean squared error of any unbiased estimator $\est{\theta}$ of $\theta$ according to:
\[ \func{MSE}{\est{\theta}} \geq \frac{1}{\func{I}{\theta}}.\]
For a probability distribution $\func{Pr}{x;\theta}$, with discrete outcomes $x$ varying continuously with $\theta$, there are two ways to increase the gradient.
\begin{enumerate}
\item \textbf{Increase the amplitude}. Ensure that as $\theta$ varies, the probability changes across the full range from `0' to `1'. Experimentally, this is often achieved by removing any technical or unrelated noise from the measurement.
\item \textbf{Reduce the width/period}. Ensure a smaller change in  $\theta$ induces the probability to vary from `0' to `1'. Experimentally this is achieved by making the measurement device more responsive to $\theta$.
\end{enumerate}

Quantum mechanics provides a set of rules that govern any realizable measurement sample $\func{Pr}{x;\theta}$ and these rules dictate that the best measurements are obtained on states that respond more to the parameter. Notably, QM does not predict that \cref{eq:CFI} can be violated, even by squeezed states. Let's address the claims of the quantum squeezing community in this respect. If we take just a single particle and perform a measurement of its state in order to estimate $\theta$, the quantum squeezing community would have us believe that some fundamental noise prevents $\func{Pr}{x;\theta}$ going from zero to one, thereby reducing the gradient of the logarithm of the probability. Furthermore, when using squeezing, the amplitude of the probability function can increase by an arbitrary amount. % Fine, let's assume for the sake of argument that such a noise source exists. The quantum squeezing community then claim that using squeezing, the range of the probability function increases by a factor of $N$. 
No quantum physics is needed to realise this is impossible.\\

%The \href{https://physics.stackexchange.com/questions/274891/interpreting-the-quantum-fisher-information}{reply} to this physics stackexchange question sums it up neatly:
%\begin{quotation}
%The classical fisher information is a measure of how quickly a probability distribution changes with respect to some parameter. While the quantum fisher information is how quickly a quantum state changes with respect to some parameter.
%\end{quotation}
There is no noise in the quantum Fisher information, just as there is no noise in a pure quantum state. This becomes self-evidentiary when a quantum state is considered as a unit ray in a complex Hilbert space. From this perspective, it is nonsensical to say that some quantum states have more noise than other quantum states, or that certain quantum states are squeezed in some direction whereas other states are not. %\footnote{It takes a little bit of maths to show that if we take our definition of uncertainty to be $\unc{\theta}=\sqrt{\func{Var}{\est{\theta}}}$, then we need $\func{Pr}{x;\theta}$ and not $\ln\left[\func{Pr}{x;\theta}\right]$ to increase by a factor of $N$ in order to reduce $\unc{\theta}$ by a factor of $N$.}.

\section*{An extended summary}

If you want to compare the precision of two measurement devices, one method is to break-down each device into its individual components and analyse the precision each component provides. If each component of the first device provides no better precision than each component of the second, it is possible to conclude that the first device cannot outperform the second. For rigour, two additional points are needed:
\begin{enumerate}
\item\label{item:point1} A check that the second device has at least as many components as the first (assuming more components improve precision).
\item\label{item:point2} An assumption that the components are independent. 
\end{enumerate}

This paper performs such an analysis, under the proviso that each component provides one measurement datum -- i.e. is readout once. The mathematical content is simply to analyse the optimum precision per unit time that quantum mechanics predicts an individual component can achieve.\footnote{To be precise, I analyse the maximum \textit{information} each individual component provides per unit time -- from a single measurement.} The physics is to associate an individual component of a measurement device with a single \hyperlink{non-separable}{non-separable} state-vector. As entangled particles form non-separable states, this analysis of the measurement precision allows for quantum correlations. However, with assumption\,(\ref{item:point2}) above, these are the only form of correlations present.

To avoid confusion, I emphasise a key point to this analysis. In a device composed of $N$ particles, we allow that some number of particles $N_E \leq N$ are entangled with each other. These $N_E$ particles form a \textit{single} \hyperlink{non-separable}{non-separable} state-vector and are associated with a single component of the measurement device. %Take for example, a device composed of $N = 6$ atoms where pairs of atoms are entangled in a Bell state, i.e. $N_E = 2$. This device has three independent components, each described by a single non-separable state-vector, namely a 2-atom entangled state. With no entanglement, the device has six independent components, whereas if all atoms are entangled then the device contains just a single component. 
In particular, in satisfying point\,(\ref{item:point1}), our proof makes explicit use of the fact that entanglement reduces the number of independent components in a device and in this context, quantum correlations are disadvantageous; unless this loss in components if offset by a gain in precision.%, specifically we show that entanglement reduces the number of independent measurement results thereby worsening the measurement precision. How is this? this seems to be the sticking point in the quantum community. Nobody seems to be willing to accept that such a thing is possible, even when presented with a clear mathematical argument. The argument follows:

Using standard quantum mechanics, it is straight-forward to bound the amount of information per unit time, that a single state-vector can provide on a signal $\theta$. The amount of information (and the measurement precision) just depends on the state-vector response to $\theta$. This is a well-known result, often attributed to Helstrom \cite{Helstrom1976} and Holevo \cite{Holevo1982} (arguably also Mandelstam and Tamm \cite{Mandelstam1945}) and widely used in quantum metrology literature \cite{Braunstein1994, Braunstein1996, Pezze2018, Pang2014, Pang2017, Jones2010, Zwierz2010, Escher2011, Hall2012b, Gefen2017, Giovannetti2006, Zwierz2012, Giovannetti2011, Boixo2007, Tsang2012, Jarzyna2015, Yuan2015, Gorecki2020, Len2022}. In terms of the mathematics and physics there is nothing even mildly controversial about what I am saying here. Note, we don't need to worry about the state-vector length -- entangled or unentangled -- any pure state has unit length.

I then make the observation that in the quantum squeezing community, people claim to improve the precision of their measurement device (beyond this limit) without improving the state response to the signal. In fact, in a squeezed device, the response to the signal of a single non-separable squeezed state is often described as being the same as that of a single particle. In case you think that I am misrepresenting the position of the squeezing community, please look through the papers for yourself%\footnote{In spin-squeezing it is actually very hard to find a mathematical definition of the squeezed states. Even the review articles do not give an example of a state vector that is a squeezed state. Have a look through \cite{Gross2012, Pezze2018}, you will find a definition of a coherent spin state as the tensor product of $N$ qubits in pure states pointing along the same mean spin direction, but there is no example of any state that could qualify as a squeezed state. In \cite{Ma2011} the first explicit example of a spin-squeezed state appears in Eq.\,(283)}
. The review papers included as Refs.\,\cite{Ma2011, Gross2012, Ludlow2015, Pezze2018} (also the references within \cite{Ma2011}), and Refs.\,\cite{Caves1981, Walls1983, Wodkiewicz1985, Wu1986, Slusher1987, Xiao1987, Polzik1992, Wineland1992, Kitagawa1993, Wineland1994, Sanders1995, Kuzmich1998, Soerensen1998, Brif1999, Kuzmich2000, Meyer2001, Orzel2001, Andre2002, Esteve2008, Appel2009, Eberle2010, Gross2010, Koschorreck2010, Koschorreck2010a, Leroux2010, Leroux2010a, LouchetChauvet2010, Riedel2010, SchleierSmith2010, Wasilewski2010, Abadie2011, Luecke2011, Sewell2012, Aasi2013, Taylor2013, Bohnet2014, Muessel2014, Strobel2014, Bohnet2016, Cox2016, Davis2016, Hosten2016a, Hosten2016, Kruse2016, Linnemann2016, Macri2016, Polzik2016, Tse2019, Braverman2019, Bao2020, Malia2020, PedrozoPenafiel2020, Schulte2020, Casacio2021, Gilmore2021, Colombo2022, Greve2022, Malia2022, Bornet2023, Eckner2023, Franke2023, Ganapathy2023, Jia2024} are a good starting point. There is a mathematical and logical contradiction between these two statements -- they cannot both be true. We cannot simultaneously believe that: \textit{a}) the precision of any single non-separable state-vector is bound by the state response to a parameter, and \textit{b}) squeezed states can improve the measurement precision beyond this bound without increasing the state response.\\

This contradiction implies that one of, statement \textit{a)} or \textit{b)} above are incorrect. In the main text, we show the squeezing community is in error.\\%\footnote{Honestly, the papers on quantum squeezing are not very clear, making it extremely difficult to find a good explanation or a consistent claim that could be discredited. I have done my best to provide an accurate summary here. Of course the reason why it is hard to find a self-consistent explanation of quantum squeezing that one could clearly refute is simple -- all of the papers contain logical contradictions. It goes to follow that if the papers were clear, than this contradiction would be obvious.}.\\

The above clearly highlights a specific contradiction promoted by the squeezing community, and equips any critical observer with a simple question to be addressed before accepting claims of an improved measurement precision. It can take the following form:
\begin{quotation}
Take just a single, \hyperlink{non-separable}{non-separable} squeezed-state in your device. Is the response of this state to the signal greater than that of a single particle?\footnote{Most preferably, write down a bound on the precision in terms of the eigenstates of the Hamiltonian with maximum and minimum eigenvalues as in \cite{Pezze2018, Pang2014, Pang2017, Braunstein1996, Jones2010, Zwierz2010, Escher2011, Hall2012b, Gefen2017, Giovannetti2006, Zwierz2012, Giovannetti2011, Boixo2007, Tsang2012, Jarzyna2015, Yuan2015, Gorecki2020, Len2022}.}
\end{quotation}

Once the state response is defined, then assuming the answer is consistent with the postulates of quantum mechanics, so is the optimal measurement precision. Furthermore, as the state response bounds the precision, any experimental evidence of an improved precision must include demonstration an enhanced sensor response. The squeezing community present an answer that contradicts this bound. The state response of the squeezed state is not increased! In squeezing, the 
\emph{only} experimental evidence for an enhanced precision is of the form of reduced noise. This is a classic example of how our cognitive biases can be convinced by incorrect analysis. Because we expect that reducing noise makes a measurement more precise we stop to ask whether this comes at the cost of the signal.% Often this contradiction is obscured by incorrect but convincing analysis that exploits one or more cognitive biases. A classic is using one or more positing some fundamental noise intrinsic to quantum states that can be reduced, and then only on the noise whilst neglecting how the signal changes. % by comparing the precision of the ensemble to a $\sqrt{N}$ scaling that is worse than the absolute bound, and forgoing analysis of the single particle precision, this is coupled 
%Because the uncertainty is usually presented after a series of mathematical steps are performed, and these steps are not clearly documented, before accepting claims of the entire device, I encourage anyone to first ask for a comparison of the precision provided by individual state-vectors. 

Although the logical inconsistencies are most obvious when analysing just a single \hyperlink{non-separable}{non-separable} state-vector, the arguments can be extended to devices containing more than one state-vector. To prove that a device constructed from $N$ particles and using squeezing can not provide more information per unit time than possible from an ideal device with $N$ independent particles (this is known as the standard quantum limit), I note:
\begin{itemize}
\item For an ensemble containing $N > 1$ particles, if the ensemble is entangled to any degree, it must contain less indivisible state-vectors than an unentangled ensemble.
\end{itemize}
I.e. the entangled ensemble has less components as per (\ref{item:point1}) above. That's it. Much of the main text discusses experiments with spins, but the analysis is general and also applies to measurement devices using photons, i.e. optical interferometers such as LIGO.

If I could pinpoint why people have such strong resistance to these statements, it seems to be because there is a vast body of theoretical and experimental work claiming the contrary. If you are at that point, then I ask you first to go through the arguments that I am making and look for an error. If none is forthcoming, then you might consider reassessing the analysis presented by the rest of the scientific community (see the \hyperlink{FAQs}{FAQ} section and Refs.\,\cite{McGuinness2021,McGuinness2022, McGuinness2023, McGuinness2023a} for a summary of my efforts to date). Even though the papers are difficult to follow, the mistakes are actually surprisingly obvious once you know where to look. For hints on where to start, I have \href{https://pubpeer.com/publications/E11048DC489BB4FBDDA5C86D2C9D27}{published} public \href{https://pubpeer.com/publications/F46AB6EE50BFAA2F2219C78E6EB054}{comments} on PubPeer. You could also try contacting the authors to get an answer to the question posed above...

\section*{Peer review statement}
%\subsection*{For the non-experts and those who don't want to make an effort}
I acknowledge that the arguments presented here are too technical for many people to follow. Even for those with the requisite background, substantial effort is required to understand these claims and uncover the errors in individual papers. % Most people seem to be unwilling to do this. \textcolor{blue}{Need a different way to convince people.}
I would like other people to check my work for rigour and mistakes, beyond that I  want to demonstrate to non-experts that this has been done. I do not think that the peer-review system performs those two tasks very well (see Refs.\,\cite{McGuinness2022, McGuinness2023, McGuinness2023a}). For those two goals, a much more effective strategy is to financially incentivise people to perform a thorough peer-review, i.e. offer a financial reward to anyone who finds an error. It is important for this to be made public, both to ensure that I am being true to my word and so that anybody can check the status of prize and read through any counterarguments. A great resource for such a public debate is PubPeer.
%Why do I refer to quantum technologies in general here, instead of just quantum squeezing? Because, this work was inspired by the following conjecture: the amount of information that can be extracted from any quantum system per unit time just depends on the number of non-separable states \cite{McGuinness2021a}. entangled states cannot outperform 

This paper comes with a US\$10,000 peer-review prize for finding any errors which directly invalidate the following statement\footnote{Payable once, to the first person.}
\begin{quotation}
If a single \hyperlink{non-separable}{non-separable} state does not respond more to a parameter $\theta$ than a spin-$1/2$ particle then the axioms of quantum mechanics directly imply this state has no more quantum Fisher information on $\theta$ than the single particle state.
\end{quotation}
Here the response of a state $\ket{\psi}$ is characterized by the change in Hilbert space angle of $\ket{\psi}$ for an infinitesimal change in $\theta$. It then follows that such a state provides no more classical Fisher information on $\theta$ than the optimal measurement on a single particle state.
%We use the following broad definition of a squeezed state -- any non-separable state that does not respond greater to a signal than a single particle state.

The prize can be collected by posting a public comment on this paper in PubPeer (it can be done anonymously, but not if you want the prize). If you are a non-expert who is interested in quantum technology, I encourage to check PubPeer whether anyone has claimed the prize. The longer it goes unclaimed, the more your confidence in the potential of quantum technologies should reduce. Note that in Refs.\,\cite{McGuinness2021,McGuinness2022} I have also offered a prize of US\$10,000 for the first experimental demonstration of a precision beyond the standard quantum limit, and an additional prize of US\$10,000 for the first demonstration using $N$-partite entanglement that surpasses the single particle limit (in defining the resources, the measurement time and number of particles should at the least include all unitary evolution). I have emailed the authors of several papers \cite{Colombo2022, Greve2022, Malia2022, Eckner2023} claiming to surpass the SQL and offered the prize if they send me the data substantiating their claims. Despite this, the prize has remained unclaimed after several years.

\newpage

\section*{Introduction}
%We first give a non-rigorous discussion of measurement precision in the context of spin-1/2 particles and outline arguments towards the proof. % a layperson explanation of the proof.\\
Consider a sensing device composed of $N$ spin-$1/2$ particles,\footnote{Note, these results apply to photon interferometry and are discussed in this context in Appendix\,\ref{App:History}.} which is used to measure the value of some unknown parameter $\theta$. %Typical scenarios include measuring the amplitude or direction of a magnetic field, frequency estimation or phase estimation. 
Taking just one spin and using the spin direction $\myvec{S}$ of this particle as a meter, then by error propagation any estimate $\est{\theta}$ of the value of $\theta$ has an uncertainty:
\be
\Delta \est{\theta} \geq \left| \frac{\partial \myvec{S}}{\partial \theta} \right|^{-1} \Delta \myvec{S}, % + \mathrm{Bias},
\label{eq:SNR}
\ee
where $\Delta \myvec{S}$ is the uncertainty in determining the spin direction%\footnote{\textcolor{blue}{We are going to take a classical parameter estimation approach here. By this we mean that the parameter $\theta$ has a defined, fixed value which we want to estimate. Then, more fully, we can include the possibility of a finite bias $\epsilon$ and need to ensure that the gradient is taken at the actual value of $\theta$. average over $\func{Pr}{\theta = \Theta}$, the \emph{a priori} probability distribution of $\theta$, to obtain the expected uncertainty: $\langle \Delta \est{\theta}\rangle \geq \left[\int_{\theta \in \Omega} \func{Pr}{\theta = \Theta}\left( \left| \frac{\partial \myvec{S}}{\partial \theta} \right|^{-2} \left(\Delta \myvec{S}\right)^2+\epsilon^2\right) \mathrm{d}\theta \right]^{1/2}$, where $\theta$ can take on values in $\Omega$ and the integral should be replaced by a sum for a discrete probability distribution. We need to include an accurate defintion of uncertainty here -- as the mean squared error in the estimator, and even generalise it to general distance metrics.}}
. The derivative term is often called the measurement signal, and $\Delta \myvec{S}$ %$(\Delta \myvec{S})^{-1}$ 
the measurement noise,\footnote{Incorrectly so, it is the measurement uncertainty not noise.} so that the `signal-to-noise ratio' determines the measurement precision. With $N$ identical and independent spins, in general one cannot do better than
\be
\Delta \est{\theta} \geq \Delta \est{\theta}_1 / \sqrt{N}, \label{eq:SQL}
\ee
where $\Delta \est{\theta}_1$ bounds the uncertainty using a single particle. Note, for Eq.\,\eqref{eq:SQL} to hold, $\Delta \est{\theta}_1$ must represent the best possible uncertainty that can be attained with a single particle, otherwise one could beat the limit simply by improving $\Delta \est{\theta}_1$. Furthermore, in general $\Delta \est{\theta}_1$ is a function of time since the spin response is given by unitary evolution and with more time one can increase the spin response or perform more measurements to improve the uncertainty. Under this definition, Eq.\,\eqref{eq:SQL} is called the standard quantum limit (SQL) and it sets an uncertainty bound per unit time that is impossible to surpass with a given number of identical and independent spins.

Two approaches have been proposed to overcoming the SQL using entanglement, either make the spin response greater, or reduce the uncertainty in measuring the spin direction \cite{Gross2012, Ma2011, Ludlow2015, Pezze2018}. The first makes use of entangled NOON, CAT or GHZ states so that, in theory: $\frac{\partial \myvec{S}_N(t)}{\partial \theta} = N \frac{\partial \myvec{S}_1(t)}{\partial \theta}$, whilst $\Delta \myvec{S}_N = \Delta \myvec{S}_1$, where the subscript denotes the number of spins in the sensor \cite{Pezze2018, Gross2012, Ludlow2015, Bollinger1996, Dowling1998, Childs2000, Campos2003, Mitchell2004, Kok2004, Leibfried2004, Nagata2007, Resch2007, Berry2009, Dorner2009, Escher2011, Giovannetti2011, Daryanoosh2018, Shaniv2018, Xie2021}. In effect a single spin with greater magnetic moment is created such that the response of the entangled device is $N$-fold greater than that of a single spin. It is worth noting here, that $\Delta \myvec{S}$ is the result of sampling from an unknown probability distribution and since a measurement of both a single entangled state and a single spin provide only one sample, they have the same uncertainty. We will not discuss this approach except to note that when the resources required to generate entanglement are fully accounted for, no improvement over the SQL or $\Delta \est{\theta}_1$ has been demonstrated\footnote{Even though this point is adjacent to the analysis presented here, it is worth emphasising. Despite the popular narrative asserting otherwise, when the total measurement time is included no entangled state, not just squeezed states but no entangled state at all, has demonstrated a precision exceeding what one can obtain with just a single spin!} \cite{McGuinness2021}.

The second approach uses entangled squeezed states so that:\footnote{Early squeezing proposals did not require entanglement to overcome the SQL, resulting in confusion on the definition of a squeezed state and whether entanglement is required (see Appendix\,\ref{App:History}). To date, the definition of squeezing and the SQL is still not uniformly standardized, see \cite{Braginsky1996, Soerensen2001,Meyer2001,Maccone2020} for different definitions. We explicitly define the SQL as whatever one can achieve without entanglement, thus it is a moot point whether entanglement is required to surpass it.} $\frac{\partial \myvec{S}_N(t)}{\partial \theta} = \frac{\partial \myvec{S}_1(t)}{\partial \theta}$, whereas $\Delta \myvec{S}_N \approx \Delta \myvec{S}_1/N$. I.e. the spin response of the entangled device remains the same as that of a single spin but the measurement noise reduces by up to a factor of $N$ \cite{Caves1981, Ma2011, Pezze2018, Gross2012, Ludlow2015, Walls1983, Wodkiewicz1985, Wu1986, Slusher1987, Xiao1987, Polzik1992, Wineland1992, Kitagawa1993, Wineland1994, Sanders1995, Kuzmich1998, Soerensen1998, Brif1999, Kuzmich2000, Meyer2001, Esteve2008, Appel2009, Eberle2010, Gross2010, Koschorreck2010, Koschorreck2010a, Leroux2010, Leroux2010a, LouchetChauvet2010, SchleierSmith2010, Wasilewski2010, Luecke2011, Sewell2012, Aasi2013, Taylor2013, Bohnet2014, Muessel2014, Strobel2014, Kruse2016, Polzik2016, Cox2016, Davis2016, Bohnet2016, Hosten2016, Linnemann2016, Macri2016, Tse2019, Braverman2019, Schulte2020, Malia2020, Bao2020, PedrozoPenafiel2020, Casacio2021, Gilmore2021, Greve2022, Malia2022}. We can already see that there is a conflict between our explanation of the origin of the measurement uncertainty and what is observed with squeezed states. Shouldn't the uncertainty in estimating the direction of a single squeezed state be the same as for a single spin? They both correspond to a single sample from a probability distribution. Results in quantum tomography actually show that the uncertainty (error in state estimation) increases with Hilbert space dimension, so it should be harder to determine the direction of squeezed states \cite{ODonnell2016, Aaronson2018}. Where does this reduced uncertainty come from? Either our layperson description is wrong or squeezed states do not perform as described.

It turns out that the above introduced a subtle error by conflating the measurement noise with uncertainty. Did you spot it?\footnote{My apologies for perpetuating this error and changing between `uncertainty' and `noise', they do not mean the same thing. I would like to consistently use the correct terminology - uncertainty, but spin squeezing papers conflate terms -- switching between noise and uncertainty -- and I am trying to present their claims. This conflation of terms is the crux of the issue and I encourage the reader to take careful note. Returning to the start of this paragraph, $\Delta \myvec{S}_N = \Delta \myvec{S}_1/N$ does not mean the measurement noise reduces by a factor of $N$, rather the uncertainty in estimating the spin direction of $\myvec{S}_N$, is $N$ times lower than estimating the direction of a single particle $\myvec{S}_1$. Think about the implication -- given a single squeezed state pointing in the same direction as a single spin, apparently one can estimate its direction $N$ times better than the single spin. What squeezing experiments show instead is a reduction of noise, this task is much easier to achieve than reducing uncertainty.} We now show that, if entanglement provides any benefit over the SQL, then it must come about from increasing the sensor response to the signal and not through reduced noise.\\

\section*{Noise independent quantum precision bound}% Quantum precision bound only depends on detector response to signal.}
For rigour we make two adjustments to our language. First, rather than talk about spin direction, we address the underlying mathematical object -- the state-vector. To show that the uncertainty bound obtained from an ensemble containing entangled squeezed states is worse than that of an unentangled ensemble, we use a counting argument based on the number of indivisible state-vectors (i.e. a state-vector that cannot be further factored into \hyperlink{non-separable}{non-separable} states) in the ensemble. Henceforth, a state-vector, $\ket{\psi}$ always refers to this basic unit we are dealing with -- an indivisible state. A single state-vector is never used to describe the quantum state of a separable ensemble, instead we keep note of the number of copies of each state. Secondly, for technical reasons and conciseness we avoid quantitatively defining uncertainty (see the \hyperlink{FAQs}{FAQ} section), instead we define the (Fisher) information on $\theta$ denoted $\func{I}{\theta,t}$, provided by a given state, or more precisely, measurement of that state \cite{Braunstein1994}:
\be
\func{I}{\theta,t} := \int_{X} \mathrm{d}x \frac{1}{\func{Pr}{x;\theta,t}}\left(\frac{\partial \func{Pr}{x;\theta,t}}{\partial \theta}\right)^2, \quad \func{I}{\theta,t} := \sum_{i=1}^R \frac{1}{\func{Pr}{x_i;\theta,t}}\left(\frac{\partial \func{Pr}{x_i;\theta,t}}{\partial \theta}\right)^2
\label{eq:FI}
\ee
where $\func{Pr}{x;\theta,t}$ is the probability to obtain the measurement result $x$ at time $t$ parametrized by $\theta$ and the (LHS) RHS assumes the measurement outcomes are (continuous taking on values in $X$) discrete with $R$ possibilities. We note that for any measurement, as long as the estimation uncertainty $\Delta \est{\theta}$ is a non-increasing function of $\func{I}{\theta,t}$, we can compare uncertainties and say one is greater than or equal to the other by relating the Fisher information. This is sufficient for our purposes.\\

\emph{Key requirement:} The central claim in squeezing enhanced metrology is that squeezed states have an improved intrinsic noise (uncertainty) compared to a single spin. It is clear that the response to $\theta$ of the squeezed state, quantified by $\lVert\frac{\mathrm{d}\ket{\psi(\theta,t)}}{\mathrm{d} \theta}\rVert$, is not greater than that of a single spin. One reason this must be the case is that if the noise can be reduced by a factor of $N$, then any further improvement in the sensor response would violate the Heisenberg limit. See \hyperlink{FAQs}{FAQ} \ref{FAQ:2} for a list of experiments in squeezing that confirm this observation.\\% Our proof directly addresses this claim, we show an information bound that just depends on the state response to $\theta$, therefore, if the state response is not increased, the uncertainty cannot be reduced beyond the SQL.

Finally, we only establish a bound on the information provided by a single measurement of the ensemble. While this allows a direct like-for-like comparison between measurement devices, it is important to note that this is not how measurement sensitivity is usually reported. Often, when reporting squeezing enhancements, comparison is made to a time-averaged limit with $\mathrm{Hz}^{-1/2}$ uncertainty improvement. Further assumptions for the proof are provided in Appendix\,\ref{App:Assumptions}.\\
%\begin{itemize}
%\item A single measurement of each particle in the ensemble.
%\item Where the path parameterized by $\theta$ is smooth. I.e. the probability distribution of $\theta$ is smooth.
%\item $\left(\Delta \est{\theta}\right)^2$ denotes the expected mean squared error of the estimator, thus $\Delta \est{\theta}$ is the positive square-root of the expected mean squared error (expected root-mean-square error) of the estimator.
%\item We are calculating the information gain, uncertainty reduction. Or something proportional to this.
%\end{itemize}

%Although it appears like a restrictive list, what types of experiments does this cover?
%Ramsey interferometry, Mach-Zehnder interferometry. Most experimental demonstrations - B field estimation, frequency estimation, phase estimation, operating regime of LIGO (assuming short measurement time so that the gravitational wave signal does not change much).

\emph{Counting approach:} Assuming we can prove --
\begin{quotation}
\hypertarget{S1}{Statement 1}: A single squeezed state does not provide fundamentally more information on $\theta$, per unit time, than a single spin.
\end{quotation}
-- it follows that the information bound on $\theta$ for any $N$ spin ensemble containing $M$ squeezed states is lower than an ensemble of $N$ unentangled spins. The reason being that a squeezed ensemble can be separated into a maximum of $M$ independent state-vectors where $M < N$ (this follows from the very definition of entanglement). Assuming these states are independent,\footnote{This assumption just ensures that any correlations come solely from entanglement.} the information provided by the squeezed ensemble is $M$ times that of a single squeezed state and is therefore less than the information provided by the unentangled ensemble. In fact, this counting argument shows that increasing squeezing leads to a worse uncertainty bound since there are less \hyperlink{non-separable}{non-separable} states to average over. Meaning that, if the uncertainty provided by an ensemble containing squeezed states ever surpasses the single particle bound and if the degree of squeezing/entanglement is continuously increased, then at some point the uncertainty must get worse. And the converse: if the measurement uncertainty always improves for arbitrary amounts of squeezing, then the uncertainty never surpasses the single particle bound.

Many mathematical statements equivalent to \hyperlink{S1}{Statement 1} (excluding the counting argument) have been published. For example, Wootters showed in 1981 that $\func{I}{\theta,t}$ can be interpreted as a distance metric over quantum states \cite{Wootters1981}, i.e. the information on $\theta$ increases with the distance between the states $\ket{\psi(\theta_1)}, \ket{\psi(\theta_2)}$ in the limit $\theta_2 \to \theta_1$. A direct implication is that if the path mapped out by unitary evolution is the same for two states then so is the information. This is equivalent to saying that the uncertainty depends only on the sensor response to $\theta$ \cite{Ou1997, Childs2000, Giovannetti2004, Zwierz2012, Pang2014, Yuan2015, Pang2017, Gorecki2020}, therefore states with no enhanced response provide no fundamental sensing advantage. Giovannetti, Lloyd and Maccone proved that (for a restricted class of Hamiltonians) entanglement cannot provide any benefit at the measurement stage \cite{Giovannetti2006} and therefore must be utilized during state evolution. Braunstein, Caves and Milburn provided most of the mathematical content of the proof (reformulating the bounds derived by Helstrom \cite{Helstrom1976} and Holevo \cite{Holevo1982}) by showing that for pure states, $\func{I}{\theta,t}$ is given solely by the gradient of the state-vector with respect to $\theta$, and does not depend on any intrinsic uncertainty of this state \cite{Braunstein1996}. Here we detail the arguments in full.

\section*{Proof - Time-independent multiplicative Hamiltonian: $\Op{H}(\theta) = \theta \cdot \Op{H}$}
Denote a spin-squeezed state as $\ket{\psi_{SS}(\theta,t)}$, and the state of a single spin as $\ket{\psi_1(\theta,t)}$. Denote the maximum information on $\theta$ that can be provided by any measurement on $\ket{\psi_{SS}(\theta,t)}$, $\ket{\psi_1(\theta,t)}$ as $\funcwrt{I}{\theta, t}{SS}$ and $\funcwrt{I}{\theta, t}{1}$ respectively, then we have:\\

\textbf{Claim:} If $\left(\frac{\mathrm{d}\bra{\psi_{SS}(\theta,t)}}{\mathrm{d} \theta}\right)\left(\frac{\mathrm{d}\ket{\psi_{SS}(\theta,t)}}{\mathrm{d} \theta}\right) = \left(\frac{\mathrm{d}\bra{\psi_1(\theta,t)}}{\mathrm{d} \theta}\right)\left(\frac{\mathrm{d}\ket{\psi_1(\theta,t)}}{\mathrm{d} \theta}\right)$ then $\funcwrt{I}{\theta,t}{SS} \leq \funcwrt{I}{\theta,t}{1}$.\\

\textbf{Physical interpretation:} Squeezed states claim to surpass the SQL by reducing the uncertainty associated with the state, and not by increasing the response of the state to the signal (i.e. the derivative). To refute this claim we need to show that if the gradient of the squeezed state with respect to the signal is the same as that of a single spin, then the information bound on the squeezed state is less than or equal to that of a single spin. As the derivative of a state-vector is also a state-vector, we can't say one state-vector is greater than another, i.e. $\frac{\mathrm{d}\ket{\psi_1}}{\mathrm{d}\theta} > \frac{\mathrm{d}\ket{\psi_2}}{\mathrm{d}\theta}$ is not a mathematical statement.\footnote{The vectors actually exist in different Hilbert spaces.} To obtain a non-negative real number, we take the inner-product. Since state-vectors are normalised, this operation returns the magnitude of the derivative.\\

\begin{proof}\hypertarget{proof}{} Braunstein, Caves and Milburn showed that for a pure state \cite{Braunstein1996}:
\be
\func{I}{\theta,t} \leq \func{QFI}{\theta,t} = 4\left[\left(\frac{\mathrm{d}\bra{\psi(\theta,t)}}{\mathrm{d} \theta}\right)\left(\frac{\mathrm{d}\ket{\psi(\theta,t)}}{\mathrm{d} \theta}\right) - \left|\bra{\psi(\theta,t)}\left(\frac{\mathrm{d}\ket{\psi(\theta,t)}}{\mathrm{d} \theta}\right)\right|^2\right].
\label{eq:BCM}
\ee
Working in the Schr\"odinger picture where time-evolution is carried by quantum states, an initial state $\ket{\psi_0}$ evolves in response to the Hamiltonian $\Op{H}(\theta,t) = \theta \cdot \Op{H}$, according to: $\ket{\psi_0} \rightarrow \Op{U}(\theta, t)\ket{\psi_0} \equiv \ket{\psi(\theta,t)}$, where $\Op{U}(\theta,t) = \func{Exp}{-\iu \theta t \Op{H} /\hbar}$. Writing $\ket{\psi_0}$ in terms of the $K$ eigenstates of $\Op{H}$, denoted by their eigenvalues $\ket{\psi_{E_k}}$ with complex amplitude $\alpha_k$: $\ket{\psi_0} = \sum_{k = 1}^K \alpha_k \ket{\psi_{E_k}}$, we have:
\[ \ket{\psi(\theta,t)} = \sum_{k = 1}^K \func{Exp}{-\iu \theta t E_k/\hbar}\alpha_k \ket{\psi_{E_k}}, \]
where the derivative of this state with respect to $\theta$ is:
\[ \frac{\mathrm{d}}{\mathrm{d}\theta}\ket{\psi(\theta,t)} = \sum_{k = 1}^K (-\iu t E_k/\hbar) \func{Exp}{-\iu \theta t E_k/\hbar}\alpha_k \ket{\psi_{E_k}}.\]
We first derive the maximum information a single spin can provide. For a spin-1/2 with two eigenstates and denoting the eigenvalues of $\Op{H}$ as $\pm \frac{\gamma}{2}$, we have:
\[ \left(\frac{\mathrm{d}\bra{\psi_1(\theta,t)}}{\mathrm{d} \theta}\right)\left(\frac{\mathrm{d}\ket{\psi_1(\theta,t)}}{\mathrm{d} \theta}\right) = \left(\frac{t \gamma}{2\hbar}\right)^2 \left[ |\alpha_1|^2 + |\alpha_2|^2 \right], \]
and
\[ \left|\bra{\psi_1(\theta,t)}\left(\frac{\mathrm{d}\ket{\psi_1(\theta,t)}}{\mathrm{d} \theta}\right)\right|^2 = \left(\frac{t \gamma}{2 \hbar}\right)^2 \left[ |\alpha_1|^2 - |\alpha_2|^2 \right]^2.\]
We can see that $\func{I}{\theta,t}$ is maximised by initially placing the spin in an equal superposition of eigenstates, so that $|\alpha_1|^2 - |\alpha_2|^2 = 0$. Then: 
%\be
\[\funcwrt{I}{\theta,t}{1} = \left(\frac{t \gamma}{\hbar}\right)^2.\]
%\label{eq:single}
%\ee
We have shown that: $\funcwrt{I}{\theta,t}{1} = 4\left(\frac{\mathrm{d}\bra{\psi_1(\theta,t)}}{\mathrm{d} \theta}\right)\left(\frac{\mathrm{d}\ket{\psi_1(\theta,t)}}{\mathrm{d} \theta}\right)$ and reproduced the well-known Heisenberg limit for a single spin \cite{Bollinger1996, Giovannetti2004, Pang2014}. Inserting into Eq.\,\eqref{eq:BCM}, the equality stated in the claim, the maximum information of the squeezed state is:
\[%\be
\funcwrt{I}{\theta,t}{SS} = \funcwrt{I}{\theta,t}{1} - 4\left|\bra{\psi_{SS}(\theta,t)}\left(\frac{\mathrm{d}\ket{\psi_{SS}(\theta,t)}}{\mathrm{d} \theta}\right)\right|^2.
\]%\label{eq:squeeze} \ee
As the second term is a non-negative number, $\funcwrt{I}{\theta,t}{SS} \leq \funcwrt{I}{\theta,t}{1}$ and the proof is complete.
\end{proof}

In the above analysis, it may seem like we only consider quantum states and not measurements, however Eq.\,\eqref{eq:BCM} implicitly contains the Born rule. In particular, $\func{I}{\theta,t}$ is a projection of $\frac{\mathrm{d}\ket{\psi(\theta,t)}}{\mathrm{d} \theta}$ meaning that (for this Hamiltonian) the optimal measurement basis is a projection perpendicular to the eigenstates\footnote{Here perpendicular means at an angle of $90^{\circ}$, not anti-parallel.} \cite{Braunstein1996, Childs2000, Giovannetti2004, Pang2017}. Explicitly, for $\Op{H} = \gamma \Op{\sigma}_z/2$ where $\Op{\sigma}_z$ is the Pauli-$z$ matrix, the measurement associated with $\funcwrt{I}{\theta,t}{1}$ is a projective measurement in the $x-y$ plane. Considering a measurement space with just two discrete outcomes, we can denote the measurement results as `1' and `0', thus allowing Eq.\,\eqref{eq:FI} to be expressed as:\footnote{We have used $\func{Pr}{0;\theta,t} = 1 - \func{Pr}{1;\theta,t}$, for a Bernoulli random variable taking only two values. %Note the number of elements in the basis is defined to be the dimension of the vector space
}
\be
\func{I}{\theta,t} = \left|\frac{\partial \func{Pr}{1;\theta,t}}{\partial \theta}\right|^{2} \frac{1}{\func{Pr}{1;\theta,t}\left(1-\func{Pr}{1;\theta,t}\right)}.
\label{eq:prob}
\ee
Here we can identify $\left|\frac{\partial \func{Pr}{1;\theta,t}}{\partial \theta}\right|$ and $\sqrt{\func{Pr}{1;\theta,t}\left(1-\func{Pr}{1;\theta,t}\right)}$ with the measurement signal and noise in Eq.\,\eqref{eq:SNR}, where the latter is called quantum projection noise \cite{Itano1993}. Note that the description of some quantum states as having intrinsically lower uncertainty is completely absent in this analysis, and in Eq.\,\eqref{eq:prob} the noise and signal are not independent.\footnote{We should be careful when equating terms in Eq.\,\eqref{eq:SNR} and Eq.\,\eqref{eq:prob} because they are not identical. In Eq.\,\eqref{eq:SNR} the response of the meter is independent of the measurement, whereas in Eq.\,\eqref{eq:prob}, $\func{Pr}{1;\theta,t}$ depends on the measurement.}

With the addition of the counting approach detailed earlier, we have proved that, for a measurement device comprised of $N$ spins, ensembles containing pure squeezed states provide a worse uncertainty bound than unentangled ensembles for estimating parameters in time-independent, multiplicative Hamiltonians. In Appendix\,\ref{App:Assumptions} we detail the assumptions in the proof and in Appendix\,\ref{App:Mixed},\,\ref{App:General} we generalise the proof to the following situations. We show that if we consider:
\begin{itemize}
\item probability mixtures of squeezed states, then the information from these mixed states cannot exceed a single spin.
\item the expected mean information, averaged over the prior probability distribution of $\theta$, then squeezed states cannot outperform a single spin.
\item a modified claim concerning the projection of the gradient $\frac{\mathrm{d} \ket{\psi}}{\mathrm{d}\theta}$ orthogonal to $\ket{\psi}$, then a squeezed state cannot cannot outperform a single spin for estimation of signals in arbitrary time-dependent Hamiltonians $\Op{H}(\theta, t)$.
\end{itemize}

\section*{Discussion}
We have proved a powerful and surprising theorem. Powerful in its generality that allows us to immediately exclude any of the methods proposed in \cite{Caves1981, Walls1983, Wodkiewicz1985, Slusher1985, Wu1986, Slusher1987, Xiao1987, Polzik1992, Wineland1992, Kitagawa1993, Wineland1994, Sanders1995, Kuzmich1998, Soerensen1998, Brif1999, Kuzmich2000, Meyer2001, Orzel2001, Andre2002, Esteve2008, Appel2009, Eberle2010, Gross2010, Koschorreck2010, Koschorreck2010a, Leroux2010, Leroux2010a, LouchetChauvet2010, Riedel2010, SchleierSmith2010, Wasilewski2010, Abadie2011, Luecke2011, Ma2011, Sewell2012, Aasi2013, Taylor2013, Bohnet2014, Muessel2014, Strobel2014, Kruse2016, Polzik2016, Cox2016, Davis2016, Bohnet2016, Hosten2016, Hosten2016a, Linnemann2016, Macri2016, Tse2019, Braverman2019, Schulte2020, Malia2020, Bao2020, PedrozoPenafiel2020, Casacio2021, Gilmore2021, Colombo2022, Greve2022, Malia2022, Eckner2023, Franke2023, Bornet2023, Braginsky1974, Braginsky1977, Unruh1979, Braginsky1980, Braginsky1996} from achieving a measurement precision beyond the SQL. Surprising in that the majority of the following experimental papers claim to demonstrate a measurement precision that violates the theorem \cite{Xiao1987, Slusher1985, Slusher1987, Polzik1992, Soerensen1998, Kuzmich2000, Meyer2001, Orzel2001, Esteve2008, Appel2009, Eberle2010, Gross2010, Koschorreck2010, Koschorreck2010a, Leroux2010, Leroux2010a, LouchetChauvet2010, Riedel2010, SchleierSmith2010, Wasilewski2010, Abadie2011, Luecke2011, Sewell2012, Aasi2013, Taylor2013, Bohnet2014, Muessel2014, Strobel2014, Bohnet2016, Cox2016, Hosten2016, Hosten2016a, Kruse2016, Linnemann2016, Braverman2019, Tse2019, Bao2020, Malia2020, PedrozoPenafiel2020, Casacio2021, Gilmore2021, Colombo2022, Greve2022, Malia2022, Eckner2023, Franke2023, Bornet2023, Ganapathy2023}. %\footnote{We need a method to physically characterise or measure the state dependence on $\theta$ since we proved that if the gradient of a spin-squeezed state is the same as that of a single state: $\left|\frac{d\ket{\psi(\theta,t)}_{SS}}{d \theta}\right| = \left|\frac{d\ket{\psi(\theta,t)}_1}{d \theta}\right|$, then both states contain the same amount of information on $\theta$. However the gradient of a state-vector is also a vector, not a non-negative real number, so to get a real number we need to take the inner product.  One might argue, isn't it tautological that if the precision is better, then by definition the gradient was increased! Perhaps people mistakenly attribute this to reduced noise and it is just an explanatory error. The problem is that it is generally obvious and easy to check the state dependence on $\theta$. Just look at the measurement results or the expected state evolution, i.e. the quantum phase accumulate per unit time. It is not the case that the phase change per unit time increases and it is mistakenly reported. Furthermore, if this information is not provided, then \emph{de facto}, sufficient information on the measurement uncertainty has not been presented. As an example take the gravitational wave detector LIGO, where squeezed light is inserted into the interferometer. The imparted phase per photon does not increase when squeezed light is inserted into the interferometer. In Ramsey interferometry with squeezed spins, it is also clear that the phase response is not increased, therefore these techniques cannot beat the SQL. A final point which can be subtle to grasp. I am not suggesting that squeezing did not improve the sensitivity of the LIGO interferometer. The results are quite clear in that regard. I am asserting that squeezing did not improve the sensitivity of the LIGO interferometer beyond a strict impossibility limit for the same number of independent input photons and measurement time.}.
This begs the question:
\begin{quotation}
Why are there so many experimental papers that seem to contradict this proof?
\end{quotation}

The answer is not that our proof is erroneous, but rather that in general, analyses of experiments either do not compare the achieved precision to the SQL or involve fundamental mistakes. Below we outline some common examples in the literature.\\% We acknowledge that the benefits of entanglement in quantum metrology are so ingrained in the consciousness that it can be difficult to overturn this conception. To help we make some ridiculous and outlandish comparisons, not just because it is fun, but because it can help to illustrate how absurd the claims are.\\

%If someone presents you with a machine that produces more output energy than input, a simple appeal to conservation of energy can be used to correct them, likewise proof of the claim, but it is somewhat non-constructive.  but sometimes it is more informative to go into detail and show exactly where the error in logic occurred.

\noindent\emph{Reasons why quantum squeezing has never surpassed the SQL:}%The reason seems to be because most experimental papers use a different definition of the SQL or incorrectly report the measurement precision.

Note how we have consistently defined the SQL as a bound on the best precision that $N$ independent and identical particles can achieve a given amount of time. It sounds silly to say, but if we instead compare to a much worse limit, there is nothing that rules out squeezing surpassing this limit. It is surprising how often this comparison is used to claim that certain experiments surpass the SQL. %Obviously if the experimental precision can be improved keeping the same number of independent particles, then the SQL has not been surpassed -- this would contradict the definition. However, w
Whenever the word \emph{entanglement} or even \emph{quantum} is used, it seems that people stop acting critically and simply accept what is presented.
%where the precision clearly does not reach the SQL. It seems that people are prepared to overlook this point whenever the word \emph{entanglement} is used.
 An analogy, let's say your mechanic promises to improve your car's fuel efficiency using some special technique, even beyond what your current car can possibly achieve. If the technique is simply to release the hand-brake, then most people would be underwhelmed. But if the mechanic calls the technique ``quantum shacky-mo'', my experience is that the scientific peer-review system would accept this without asking further questions.

\noindent\textit{A. Comparison to a different limit.}
\begin{enumerate}
\item \textbf{Comparison to an imperfect limit:} The following papers definitely seem to claim a precision per unit time that cannot be achieved using the same number of independent particles \cite{Abadie2011, Aasi2013, Taylor2013, Aasi2015, Tse2019, Casacio2021, Ganapathy2023}. It is not the case. I suggest reading through the peer review file of `quantum-enhanced' microscopy \cite{Casacio2021} where Referee 1 notes that the microscope performance is far from state-of-the-art. How do the authors improve the sensitivity? They simply align the microscope! If you can substantially improve the device performance through alignment it is clear that comparison is not to a strict impossibility limit and definitely not the same as demonstrating a measurement precision beyond the SQL. LIGO also compare to an imperfect limit, even though the sensitivity is impressive, it is far below the SQL\footnote{Up to 10 orders of magnitude.} \cite{Abadie2011, Aasi2013, Aasi2015, Tse2019, Ganapathy2023}. % If I promise to improve the performance of your car using \emph{Shacky-mo!}, and provide a demonstration where the speed of a run-down old car is improved slightly, you might wonder whether \emph{Shacky-mo!} will work on your car specifically or improve performance in general. In the context of squeezing, we can ask whether the sensitivity improvement is with respect to a fundamental limit that could not have been achieved otherwise. The answer is no\footnote{Take the work of Casacio et.\,al. \cite{Casacio2021}, where squeezed light lowers the microscope noise without increasing the signal and the authors claim to beat the SQL (they use the term shot-noise limit). In the context of LIGO \cite{Aasi2013, Tse2019} the analogy of a run-down car is less apt, since this is one of the most sensitive detectors that has ever been developed. However, one can clearly read in LIGO's 2022 \href{https://dcc-llo.ligo.org/public/0182/G2200736/001/UpdateonLVKDetectors.pdf}{update} that significant technical noise sources remain which mean that the interferometer is operating far away from the SQL. A factor of 2 -- 3 sensitivity improvement for a device operating orders of magnitude away from the SQL is not the same as demonstrating an improvement beyond this limit.}.

\item \textbf{Tautological definition of the measurement precision:} A different quantum limit is introduced. The standard quantum limit is defined as the operating precision of a measurement device \emph{as it is}, together with all of the imperfections that entails. Modifications that utilise entanglement are allowed, whilst forbidding any other changes to the device to improve it \cite{Wu1986, Xiao1987, Taylor2013}. Of course if we forbid any other changes to the measurement device then the precision cannot be improved. In these papers, the entanglement modifications also happen to involve using more photons, the trick is to carefully draw the boundary of the measurement device so these photons are excluded. % on The tricky part is to allow modifications that utilise entanglement but forbid any other changes to the device. One example of this definition is the work of \cite{Wu1986, Xiao1987}. 
%\item \textbf{Time not fully accounted for:}
\end{enumerate}
Note, in both of these cases, if the device was already operating at the SQL, and squeezing implemented whilst keeping the number of particles and time constant, then the sensitivity cannot improve.\\

A second argument the squeezing community employs is to present evidence of reduced measurement noise. There are however many trivial ways to reduce measurement noise without improving the measurement precision. One method is to `turn off' the measuring device so that it does not produce random results, this has the effect of reducing noise, but also precision. Another is to take the measurement data and scale it towards zero or simply throw away data-points. The technique used in \cite{Malia2022} is to perform one measurement followed by another that gives approximately the same result; and then to subtract the difference of these two results. It is obvious that although this strategy reduces the measurement noise, it also removes the measured signal and cannot surpass the SQL.

\noindent\textit{B. Analytic errors}
\begin{enumerate}
\item \textbf{Analysis of measurement noise not uncertainty:} Conflation of noise and uncertainty is one of the most common analytical mistakes in spin-squeezing, see e.\,g. \cite{Orzel2001, Esteve2008, Appel2009, Gross2010, Koschorreck2010, Koschorreck2010a, Leroux2010, LouchetChauvet2010, Riedel2010, SchleierSmith2010, Sewell2012, Cox2016, Hosten2016, Kruse2016, Colombo2022, Greve2022, Malia2022, Eckner2023, Franke2023, Bornet2023}. As an example, assume a projective measurement in the $x$-basis is performed on a single spin pointing in an unknown direction in the $x-y$ plane. Denote this state $\ket{a}$ and label the outcomes of the observable $\Op{S}_x = \bm{\sigma}_x/2$ as $\{0, 1\}$. The uncertainty in estimating the $x$-component of the spin direction is actually independent of the measurement variance. If ten experiments produce the following data-set: $\{1,1,1,1,1,1,1,1,1,1\}$, the SQL has not been surpassed, even though the variance is less than 1/4. Compare to the data-set obtained from a different state $\ket{b}$: \{0,1,1,1,0,0,1,1,1,0\}. Someone with knowledge of $\ket{a}$, $\ket{b}$ can prove this is not a statistically anomaly by showing $\func{Var}{\Op{S}_x}_{\ket{a}} < \func{Var}{\Op{S}_x}_{\ket{b}}$. Here $\func{Var}{\Op{S}_x}$ is the variance of the $\Op{S}_x$ operator and the subscript indicates the state that the variance is calculated with respect to. Despite this mathematical inequality, the uncertainty in estimating the $x$-component of $\ket{b}$ is the same as for $\ket{a}$. Ultimately we do not want measurements with less \emph{noise}, but measurements that provide lower \emph{uncertainty} on the spin direction.\footnote{In the literature $(\Delta \myvec{S})^2$ is often incorrectly equated with the variance of some spin operator, $\func{Var}{\Op{A}}$. These are not the same quantities! In general a collective spin observable $\tilde{\Op{S}}_z = \sum_{i=1}^N \Op{S}_z$ is measured, such that a variance $\func{Var}{\tilde{\Op{S}}_z} <|\myvec{S}|/4$ is presented evidence of an enhanced precision, where $|\myvec{S}|$ denotes the spin vector length. Note, in Eq.\,\eqref{eq:SNR}, $\Delta \myvec{S}$ is the uncertainty in estimating the spin direction, it is not the (square-root) variance of an operator, nor is it the square-root variance of some dataset.} Often prior information from state preparation (and the value of $\theta$) is used to ensure that the measurement statistics have low variance without reducing the uncertainty in estimating $\theta$.
%It is easy to perform measurements with low noise -- throw away the data. Given $N$ identical copies of a state, the noise can be reduced by a factor of $\sqrt{N}$ by measuring only one state. According to squeezing logic, this strategy surpasses the SQL. Thus we can see that some processes which increase the measurement noise actually reduce the measurement uncertainty.

\item \textbf{Replace the signal gradient with the contrast:} As we have seen and was pointed out in \cite{Itano1993}, some measurements can reduce the projection noise, but they come at a cost of reducing the signal gradient. Often this effect is not taken into account because the gradient is assumed to remain constant for all measurements and the measurement signal (c.f. the derivative in Eq.\,\eqref{eq:prob}) is replaced with a constant term that does not depend on $\theta$ or the measurement basis. A common substitution is to use the spin vector length, characterised by the `contrast', `visibility', `coherence' or `Wineland squeezing parameter' in place of the gradient \cite{Meyer2001, Esteve2008, Gross2010, Leroux2010, Leroux2010a, Riedel2010, Hosten2016, PedrozoPenafiel2020, Malia2020, Malia2022, Eckner2023}. Here, a measurement basis is chosen to reduce the measurement noise for certain values of $\theta$ without accounting for the commensurate reduction in signal. I.e. the experiment is designed so that $\func{Pr}{x;\theta,t} \rightarrow 0$, but the loss of signal $\frac{\partial \func{Pr}{x;\theta,t}}{\partial \theta} \rightarrow 0$ is not fully accounted for, and is often assumed to be constant. By operating close to a dark fringe, Advanced LIGO \cite{Aasi2015, Tse2022, Ganapathy2023} and QND proposals \cite{Braginsky1980, Kimble2001} suffer from this error.

\item \textbf{Fair sampling:} A final claim for improved precision is to design an experiment such that when data is observed it has higher sensitivity, but the fact that this comes at a cost -- a lower probability to obtain meaningful data -- is not fully accounted for \cite{Abadie2011, Aasi2013, Tse2019, Ganapathy2023}.
%\item \textbf{Incorrect definition of the SQL:} The SQL is commonly defined a time-independent measurement precision or defined with a square-root time dependence, e.g. \cite{Leroux2010a}.
\end{enumerate}

%What is going on here? Obviously I need to address that in at least some experiments, squeezing does improve the measurement precision, albeit not beyond the SQL. At the very least one should admit there is a dilemma here which should be investigated. When discussing the assumptions, an explanation for some of these observations is provided in Appendix\,\ref{App:Assumptions}).

There are valid questions on how one should interpret the presented proof that squeezed states cannot surpass the SQL. For instance it only addresses the fundamental probabilistic noise from quantum measurements,\footnote{It also neglects measurement back-action onto the signal. However back-action is minimised when using a single particle sensor.} of course other technical sources of experimental noise should be reduced. One question is to what extent this is possible, whether this noise is dependent on specific quantum states or measurements, and more broadly whether the information bound of Eq.\,\eqref{eq:BCM} is achievable? If not achievable, this may allow for some practical regime where squeezing can be of advantage. However, as we note in Appendix\,\ref{App:Assumptions} invoking this explanation is tantamount to saying quantum mechanics is incorrect. Perhaps a more nuanced interpretation is to note that no experiment can reach the precision bounds set by quantum mechanics since reaching these bounds requires perfect, instantaneous measurements with no decoherence. But you can't have it both ways. The squeezing community are not claiming that the SQL cannot be reached, indeed quite the opposite, the claim is that squeezing can surpass the SQL.

%\textit{There are valid objections to this proof. Only deals with fundamental quantum measurement noise, for other sources of course the noise should be reduced. But entanglement doesn't help with this. Perhaps there is a reason why we can't physically reach this bound for a single spin, and a much stricter limit holds for measurements of a single spin. but we can for an entangled spin. Note for a single spin the optimal measurement is also relatively easy to implement, it is just a projective measurement in the basis spanned by the superposition states. of the spin into one of the eigenstates. Perhaps, decoherence, ...}

\section*{Conclusion}
At a fundamental level, quantum mechanics dictates that measurement uncertainty is a statistical sampling phenomenon, since all measurement data are described by a probability distribution. Some analyses of measurement precision mistakenly associate the fundamental randomness of measurement observations as a noise which should be reduced. Here we have shown that approaches to improve the measurement uncertainty by reducing this noise cannot succeed. To misquote Rolf Landauer \cite{Landauer1998} -- \textit{The signal is the noise!} The only way this sampling noise can be overcome is by taking more samples, and a key point of this work is to emphasize that for any ensemble, entanglement reduces the number of samples that can be obtained. A second point introduced here is the description of a quantum state a unit length ray. This implies that the sensitivity of a quantum state is determined solely by how much the ray changes in response to changes in a parameter.

By relating metrological performance to the number of \hyperlink{non-separable}{non-separable} states we have proved that squeezed ensembles cannot outperform unentangled ensembles in sensing. The proof was inspired by the proposition of a general quantum mechanical uncertainty limit that applies to all \hyperlink{non-separable}{non-separable} state-vectors\footnote{The proposition in \cite{McGuinness2021a} is related to the Heisenberg time-energy uncertainty principle. In time $t$, the amount of information on $E$ the difference between any two eigenstates of a non-separable state-vector is bounded by $\func{I}{E,t} < (t/\hbar)^2$. If the Hamiltonian $\Op{H}$ depends on $\theta$, then $E$ can be parametrized by $\theta$.} \cite{McGuinness2021a}. Here non-separable wavefunctions are considered the fundamental information unit in computers and sensors, not particles. My position is that this information per unit time limit holds for all entangling procedures, not just squeezing, and its application extends beyond sensing to include any quantum information processing tasks, for instance quantum computation. Finally, it should be apparent that in this field of quantum physics, the peer-review process has failed (see also \cite{McGuinness2022, McGuinness2023, McGuinness2023a}) and this has led to a loss of scientific progress.

\bibliographystyle{naturemag}
\bibliography{references.bib}%{}

\newpage

\section*{Appendix}
\subsection{Presentation of \cref{eq:QFI} in the literature}\label{App:refs}
The following papers present the same analysis as \cref{eq:QFI}, but they do not make the connection I am making -- that \cref{eq:QFI} rules out quantum squeezing as a mechanism for improving the quantum Fisher information. In many of the following papers the parameter to be estimated is denoted $\varphi$.
\begin{enumerate}
\item Giovannetti, Lloyd and Maccone \cite{Giovannetti2006} ``It implies that the minimum error $\delta \varphi$ is obtained when $\Delta h$ is maximum. Hence the maximum $\Delta h$ is achieved by preparing ... equally weighted superpositions of the eigenvectors $\ket{\lambda_M}$ and $\ket{\lambda_m}$ of (the Hamiltonian) $H$ corresponding, respectively, to the maximum and minimum eigenvalues $\lambda_M$ and $\lambda_m$.'' To ``obtain a classical bound
\[ \delta \varphi \geq 1/[\sqrt{N}\left(\lambda_M - \lambda_m\right)] \quad (\text{Eq. 3}),\]
by preparing all the probes in the state $(\ket{\lambda_M} + \ket{\lambda_m})/\sqrt{2}$... On the other hand if $\ket{\psi}$ can be entangled, the maximum $\Delta h$ corresponds to a $\ket{\psi}$, which is an equally weighted superposition of the eigenvectors relative to the maximum and minimum eigenvalues of the global generator $h$, i.e., $N\lambda_M$ and $N\lambda_m$. ... gives an optimal error of
\[ \delta \varphi \geq 1/\left[N\left(\lambda_M - \lambda_m\right)\right] \quad (\text{Eqs. 4, 6, 7}).\text{"}\]
\item Zwierz, P\'erez-Delgado, and Kok \cite{Zwierz2012} ``This bound is saturated by the following family of maximally entangled states of the probe
\[ \ket{\psi} =\frac{1}{\sqrt{2}}\left(\ket{h_{\text{max}}} +e^{-\iu \varphi} \ket{h_{\text{min}}}\right), \quad (\text{Eq. 8})\]
where $\ket{h_{\text{max}}}$ and $\ket{h_{\text{min}}}$ are the eigenstates corresponding to the maximal and minimal eigenstates of $\mathcal{H}$."
\item Giovannetti, Lloyd and Maccone \cite{Giovannetti2011} ``Assume that the total number $n = \nu N$ of probes is split into $\nu >> 1$ groups of $N$ elements, where each group is prepared in the same entangled input state ... yielding
\[ \delta X \geq \frac{c}{N\sqrt{\nu}} \quad (\text{Eq. 5}) \]
where $c$ is the inverse of the largest gap in the spectrum of the generator $H$ of the transformation. For $N > 1$, equation (5) presents a $N^{1/2}$ enhancement with respect to the SQL. The term on the right can be attained by preparing the probes of each group in the entangled state $\ket{\psi} = \left(\ket{\lambda_{\text{max}}} +\ket{\lambda_{\text{min}}}\right)/\sqrt{2}$, where $\ket{\lambda_{\text{max}}}$ and $\ket{\lambda_{\text{min}}}$ are the eigenvectors associated with the maximum and minimum eigenvalues of $H$, respectively."
\item Boixo, Flammia, Caves and Geremia \cite{Boixo2007} ``When entanglement is allowed, however, one can choose the initial probe state to be the ‘cat state',
\[ \frac{1}{\sqrt{2}}\left(\ket{\lambda_M, \dotsc \lambda_M} + \ket{\ket{\lambda_m, \dotsc \lambda_m}}\right), \quad (\text{Eq. 2})\]
where for system $j$, $\ket{\lambda_M}$ ($\ket{\lambda_m}$) is the eigenstate of $h_j$ with maximum (minimum) eigenvalue $\lambda_M$ ($\lambda_m$)." 
\item Pang and Brun \cite{Pang2014} ``so the maximal quantum information is
\[ F_{\text{max}} = \left[ \lambda_{\text{max}}(h_g) - \lambda_{\text{min}}(h_g)\right]^2, \quad (\text{Eq. 11})\]
where $\lambda_{\text{max}}(h_g)$ and $\lambda_{\text{min}}(h_g)$ are the maximal and minimal eigenvalues of $h_g$, respectively.
\item Jarzyna and Demkowicz-Dobrza\'nski \cite{Jarzyna2015} ``In this case the optimal C-R bound takes the form:
\[ \Delta \varphi \geq \frac{1}{N\left(\lambda_{+} - \lambda_{-}\right)}, \quad \text{Eq. (2)}\]
with $\lambda_+$ and $\lambda_-$ being respectively the maximal and the minimal eigenvalues of $H$, while the optimal input state is defined using the corresponding eigenstates $\ket{\psi} = \left(\ket{+}^{\otimes N} + \ket{-}^{\otimes N}\right)/\sqrt{2}$."
\item The introduction of Yuan and Fung \cite{Yuan2015} makes the point very clearly. ``This problem is well studied in quantum metrology when the Hamiltonian is in the multiplicative form $H(x) = xH$; in this case it is known that the optimal strategy is to prepare the initial state as $\left(\ket{\lambda_{\text{max}}} +\ket{\lambda_{\text{min}}}\right)/\sqrt{2}$ where $\ket{\lambda_{\text{max (min)}}}$ is the eigenvector of $H$ for the maximum (minimum) eigenvalue; the standard deviation of the optimal unbiased estimator of $x$ then scales as $1/\sqrt{n J}$; here, $n$ is the number of times that the process is repeated and $J = \left(\lambda_{\text{max}} - \lambda_{\text{min}}\right)^2T^2$ is the maximal quantum Fisher information."
\item Pang and Jordan perform analysis for a general time-dependent Hamiltonian \cite{Pang2017} (this is the only fully general analysis that I know of) ``we finally arrive at the upper bound of the quantum Fisher information $I_g^{(Q)}$,
\[ I_g^{(Q)} \leq \left[\int^T_0 \left(\mu_{\text{max}}(t)-\mu_{\text{min}}(t)\right) \mathrm{d}t\right]^2. \quad (\text{Eq. 7})\]
It shows that the upper bound of the quantum Fisher information $I_g^{(Q)}$ is determined by the integral of the gap between the maximum and minimum eigenvalues of $\partial_g H_g(t)$ from time $0$ to $T$.''\footnote{Pang and Jordan define the maximum and minimum eigenvalues of $\partial_g H_g(t)$ at time $t$ as $\mu_{\text{max}}(t)$, $\mu_{\text{min}}(t)$.}
\item G\'orecki, Demkowicz-Dobrza\'nski, Wiseman and Berry \cite{Gorecki2020} ``In a generalized phase estimation scenario, evolution of a probe system is given by a unitary $U_{\varphi} = \text{exp}(\iu \varphi \Lambda)$, where $\Lambda$ is an arbitrary Hermitian generator of the transformation. We allow the spectrum of $\Lambda$ to be arbitrary, apart from being bounded from above and from below by $\lambda_+$ and $\lambda_-$, respectively. Analysis of the problem using the concept of the quantum Fisher information (QFI) and the quantum Cram\'er-Rao (CR) bound leads to the following SQL and HL, respectively
\[ \Delta \varphi_{\text{SQL}} \geq \frac{1}{\sqrt{n}\left(\lambda_+ + \lambda_-\right)}, \quad \Delta \varphi_{\text{HL}} \geq \frac{1}{n\left(\lambda_+ + \lambda_-\right)}, \quad (\text{Eq. 1}) \]
\item Len, Gefen, Retzker and Kolody\'nski \cite{Len2022} ``for the unitary encoding $\psi(\theta) = U_\theta[\psi]$ the quantum fisher information just reads
\[ \mathcal{F} = \left(\lambda_{\text{max}}(h_\theta)-\lambda_{\text{min}}(h_\theta)\right)^2, \quad (\text{Eq. 4}) \]
where $h_\theta = -\iu (\partial_\theta U_\theta)U_\theta^\dagger$, $\lambda_{\text{max}}(h_\theta)$ and $\lambda_{\text{min}}(h_\theta)$ are the maximum and minimum eigenvalues of $h_\theta$, respectively, and $\mathcal{F}$ is attained by $\psi(\theta)$ being an equal superposition of the corresponding two eigenvectors."
\end{enumerate}

This analysis is so clear that I don't see how one can misinterpret it. There is one, and only one strategy to improve measurement precision. Place your sensor in an equal superposition of eigenstates with maximum and minimum eigenvalues. This is a CAT/GHZ/NOON state, and it is a two-dimensional subspace of the Hilbert space. Equivalently, the sensitivity of a quantum state is bounded by the energy difference between eigenstates, this determines how much the state responds to the parameter. Although not explicitly stated, each of these equations also includes a `no-go' theorem. If the difference between eigenvalues is below this bound then so is the quantum Fisher information. What this analysis conclusively rules out is using states with a small difference in eigenvalues to improve sensing precision. I.e. states that do not respond quicker. But that is what squeezed states claim to do! Squeezed states don't respond quicker, they don't increase the signal instead they lower the noise.

\subsection{A confusing history of early ideas to surpass the SQL}\label{App:History}
How does squeezing improve the sensitivity of LIGO? Unfortunately, the groups involved provide conflicting explanations. This is both confusing and extremely frustrating. Whereas a single, clear discussion of how a machine operates allows for criticism and discussion, when the explanation changes at each iteration that task is difficult. Like arguing with a child who constantly shifts explanations, one effect is to make LIGO impervious to scrutiny. %\footnote{In the many conversations I have had, the experts use this ambiguity to discount my critique. The counter has the  }
The following presents several of LIGO's claims.

\subsection*{Quantum non-demolition measurements -- LIGO}
One of the initial proposals to improve the sensitivity of gravitational wave detectors was to use special projective \textit{measurements} with reduced noise, and not special quantum states \cite{Braginsky1974, Braginsky1977, Unruh1979, Yuen1983, Braginsky1996}. These low noise measurements, called quantum non-demolition (QND) measurements, are performed by choosing a measurement operator so that the meter is in an eigenstate of the observable and the variance of the measurement observable goes to zero \cite{Unruh1978, Braginsky1980}. The logic is that when measuring an eigenstate the measurement outcome is perfectly deterministic and there is no associated measurement noise. Furthermore, even if the quantum state is not initially in an eigenstate, after a QND measurement it is and a subsequent measurement has low noise.

We can summarize some reasons as to why QND measurements do not surpass the SQL, in particular why the second measurement is not informative. First, the requirement of prior knowledge about the state of the meter in order to perform a QND measurement is self-defeating, we need to know \emph{a priori} that the meter is in an eigenstate and if we already know the state of the meter prior to the measurement then nothing is gained from the measurement result.\footnote{If we are being pedantic, one can learn about \emph{which} eigenstate the meter is in, and perform discrimination but that is not what these proposals suggest \cite{Braginsky1974, Braginsky1977, Unruh1979, Braginsky1996}.} In fact, if we want the noise to remain low by keeping the meter in an eigenstate after the first measurement, then by definition nothing happens to the meter's state and it does not respond to the signal?! %Also there is an inherent lack of recognition of the difference between `uncertainty' and `noise'. 
In general, QND schemes neglect to consider that the measurement signal and measurement noise are not independent, measurements with reduced noise -- where the meter is near an eigenstate of the measurement operator -- also have less signal, they do not provide lower uncertainty when estimating a parameter.

Wootters explained this precisely in 1981 \cite{Wootters1981}, showing that even in a $2^N$ dimensional Hilbert space, the optimal measurement yields a probability function of the form: $\func{Pr}{x;\theta} = \func{cos^2}{\frac{m}{2}(\theta - \theta_0)}$, where $m$ is an integer and $\theta_0$ is a constant; thus when the probability to measure an observable goes to one or zero, the gradient disappears. Moreso, Itano et. al. noted that since projective measurements are not perfect, it is preferable to choose a measurement that maximises projection noise \cite{Itano1993}. William Unruh actually used related arguments in response to initial proposals on squeezing and QND measurements, but for some reason this criticism was ignored \cite{Unruh1978}. Nor was Unruh alone in his scepticism, Robert Lynch \cite{Lynch1984} argued against Horace Yuen's QND proposal \cite{Yuen1983} and strangely enough even after proposing using squeezed vacuum to improve gravity wave detection Caves presented arguments to discount these proposals and reinstate the SQL as a fundamental bound \cite{Caves1985}. The message here is that simply reducing measurement noise does not guarantee better precision, and perhaps counter-intuitively, in general measurements which maximise (projection) noise are better (see also \hyperlink{FAQs}{FAQ} \ref{FAQ:3} on the Heisenberg uncertainty relation, where the information is optimised by maximising the variance of an observable).

Finally, Giovannetti, Lloyd and Maccone proved, for a class of Hamiltonians that apply to gravity wave detection, that entangled states and not measurements are required to surpass the SQL \cite{Giovannetti2006}. In fact no entanglement is required to realize QND measurements! A projective measurement on any state produces an eigenstate of the measuremement basis and yields zero variance results under a second projective measurement. Indeed, this observation was the cause of early confusion as to what constitutes a squeezed state and whether entanglement is required. To point out the obvious, if no entanglement is used the SQL cannot be surpassed. % One unfortunate relic of this early analysis which persists today is that the SQL is often defined as the minimum obtainable variance on one observable when the variance equals that of another conjugate observable. Often this statement is incorrectly reduced to saying that the uncertainty in measuring spin-direction is equivalent to the spin noise, giving the SQL: $\Delta \myvec{S} = \sqrt{|S|}/2$. Thus when the spin noise is $< \sqrt{|S|}/2$, the SQL for measuring spin direction has been surpassed. 

Around the time of Caves' proposal to use squeezed vacuum to improve photon interferometry \cite{Caves1981}, the emphasis became on states of the meter (i.e. quantum state-vectors) rather than measurements with reduced uncertainty. These methods were then extended to spin states in Ramsey interferometers, called spin-squeezed states \cite{Kitagawa1991, Wineland1992, Kitagawa1993}. Yet one relic\footnote{My characterisation of QND measurements as a discounted historical idea is not accurate. Carlton Caves informs me that LIGO are still working on QND measurements as outlined here \cite{Kimble2001} and recently published here \cite{Ganapathy2023, Jia2024}.} survived from these early ideas to surpass the SQL; QND type measurements are currently used in the preparation of squeezed spin states \cite{Kuzmich1998, Kuzmich2000, Appel2009,  Koschorreck2010, LouchetChauvet2010, SchleierSmith2010, Sewell2012, Bohnet2014, Cox2016, Hosten2016, Hosten2016a, Braverman2019, Malia2020, Greve2022, Malia2022}. Here one starts with an ensemble of spins in an optical cavity, initialised to $\ket{0}^{\otimes N}$ whereupon a $\pi/2$-pulse rotates $\myvec{S}$ into the $x - y$ plane. The idea is to perform a QND measurement on the ensemble to project a sub-ensemble of the spins to an eigenstate -- either $\ket{0}$ or $\ket{1}$. In general, an optical probe pulse is inserted into the cavity and the phase (frequency) shift of the output light is recorded. This is a dispersive population measurement where the shift is proportional to the number of atoms projected into $\ket{0},$ $\ket{1}$. Apparently, after the QND measurement, the ensemble is now in an entangled state with less noise and therefore improved measurement precision.

I strongly disagree with this description of the ensemble. After the QND measurement, I would describe a single state in the ensemble as being in a probability mixture of $\ket{0}, \ket{1}$ and $\ket{x}$.\footnote{With the probability function determined by the measurement strength and the outcome of the dispersive population measurement. Also the state is not exactly $\ket{x}$, the electromagnetic field of the optical probe is said to slightly rotate this state around $z$.} To refute this description and in particular to demonstrate entanglement at least one piece of information is required: a definition of the phase relationship between entangled basis states. For example, one can write an entangled Bell state as: $(\ket{00} + \ket{11})/\sqrt{2}$, and although it is generally overlooked, a well-defined phase between the states $\ket{00}$ and $\ket{11}$ is required to generate and make use of this entangled state (in this example the state can be written $(\ket{00} + e^{\iu \phi }\ket{11})/\sqrt{2}$ with $\phi = 0$). Otherwise we have a mixed state and the best we can say is that for any measurement we will record either a `0' or a `1', each with 50\% probability. None of the papers that use QND preparation of a ``squeezed" spin state define or make use of any phase relationship between the entangled basis states. It is true that knowledge of the phase of the initial $\pi/2$-pulse and the intensity the probe light is required when measuring these states, but these parameters do not determine $\phi$ (see Assumptions and \hyperlink{FAQs}{FAQ} section below).

\subsection*{Squeezing and vacuum noise in photon interferometry -- LIGO}
%In the following analysis I am going to be making several approximations, or assumptions, simplifications. The purpose of these simplifications is to allow me to address the simple question ``Can squeezing provide a fundamental improvement over non-squeezing". To do this, we want to remove all other noise sources. For example, advanced ligo used an additional LO field to produce power variations that are detected by the photodiode.... Equivalent to lock-in detection for improving the signal to noise of many devices.
%

In the early development of the theory behind optical interferometers, an initial concern was that laser power fluctuations would limit the ultimate sensitivity. In response some scientists argued that these fluctuations should be evenly distributed by the beam-splitter and get cancelled out in a detector sensitive to a relative path or power difference. The matter appeared to be settled when Carlton Caves introduced a source of noise called ``vacuum fluctuations'':
\begin{quotation}
In principle the photon-counting error has nothing to do with laser-power fluctuations, whose effect is negligible very close to a null in the fringe pattern.\cite{Caves1980}
\end{quotation}
Thenceforth, the unavoidable presence of the vacuum noise entering the beam-splitter became the dominant narrative, and indeed was the explanation taught to me as an undergraduate.

More recently the description has gone back to talking about photon shot-noise present in standard laser light sources which randomises the arrival time of photons on the detector \cite{Tse2019}. Scientists working at LIGO have \href{https://news.mit.edu/2023/ligo-surpasses-quantum-limit-1023}{spoken} about photon arrival time noise that can be reduced by making ``the photons arrive more regularly, as if the photons are holding hands rather than traveling independently''. In the last few years, the idea of ``back-action avoiding'' measurements has become in vogue. The explanation is that upon reflecting off the mirror at the end of the interferometer arm, each photon delivers momentum which pushes the mirror. As each photon measures the position of the mirror, this in turn produces back-action on the mirror position. To avoid this, two of the recent LIGO publications \cite{Ganapathy2023, Jia2024} describe using the QND measurements suggested by Braginsky and collaborators, arguing this is how the sensitivity is improved.

%However this seems to be precisely the noise mechanism that Carlton Caves was refuting in his analysis:

As the previous section debunked proposals of QND measurements, below I focus on explanations that invoke either vacuum noise or photon shot-noise, explaining why they are incorrect.

\subsection*{Carlton Caves' lasting mistake}
In seminal papers on quantum mechanical noise in an interferometer \cite{Caves1980, Caves1980a,Caves1981}, Caves proposed a fundamental noise source that ``can be attributed to vacuum (zero-point) fluctuations in the electromagnetic field, which enter the interferometer from the unused input port" \cite{Caves1981}. One typical physical description of this noise is that when an electromagnetic wave enters the interferometer at the first beam-splitter, vacuum fluctuations entering the unused port cause an unequal splitting of the electromagnetic field. Caves noted an equivalent view that treats photons as particles and attributes this imbalance to the ``random scattering of the input photons at the beam-splitter". As a result more of the input light is directed to one arm of the interferometer and less into the other, in turn affecting the output light intensity at both ports and imparting different momentum on the interferometer mirrors.
\begin{quotation}Suppose $N$ photons are incident on the beam splitter. It scatters each photon independently, thereby producing a binomial distribution of photons in each arm of the interferometer. The two binomial distributions are precisely anticorrelated (if too many photons go down one arm, too few go down the other), and this is precisely what is necessary to produce a $\sqrt{N}$ fluctuating force \cite{Caves1980}.
\end{quotation} 

 It is commonly understood that this noise cannot be removed, except through inserting squeezed ``vacuum" into the unused input port to reduce these fluctuations. Here I present a different physical description of the interferometer, one that changes the emphasis of the noise and that better predicts experimental results.

As in the main text, one can learn a lot by reducing the experiment down to its individual components and providing a physical description of this basic experiment. So what happens when a single photon enters the interferometer, whereby it reflects off the end mirrors and is detected at the unused output port (or other port)? Using a particle description of the photon, which randomly scatters at the beam-splitter, we encounter a problem. The photon only traverses one arm, thus there is no interference. No matter the path difference, upon returning to the beam-splitter the photon exits the unused port with 50\% probability. This is not what is observed experimentally. Furthermore, the idea of each photon traversing just one arm of the interferometer leads one to infer that the output signal depends on interference between \textit{separate} photons. This does not capture the basics, rather it is interference of a photon with \textit{itself} that is relevant.\footnote{More precisely it is interference between basis states of a single wavefunction, which can include more than one photon for entangled states.} The particle description is not accurate in describing this experiment. What if we take the quantum mechanical description of a photon as a wavefunction instead? Now the suggestion is that vacuum fluctuations somehow prevent one from creating an equal superposition state at the first beam-splitter. However this conflicts with experimental results -- the creation of equal superposition states is a basic concept in quantum metrology and quantum information processing. For example, the observed probability (up to experimental imperfections) to detect a photon at the unused port is given by $p = \sin^2(\phi/2)$, where $\phi$ is the relative phase difference. If vacuum fluctuations always enter the unused port to randomise the observed photon statistics, then how can one explain this observation? Take $\phi = 0$. Without inserting any squeezed vacuum into the unused port, one can predict the experimental outcome with remarkable accuracy. Fringe visibilities in excess of 99\% have been demonstrated for single photon interference \cite{Jacques2008, Nardelli2026}. This should not be possible if vacuum fluctuations of order $\sqrt{N}$ are a real and fundamental source of noise that cannot be removed.

The explanation of a photon imbalance in each arm of the beam-splitter is incorrect, and anyone with a firm understanding of quantum physics and who is motivated to question the description can see why. A different analysis to Caves' is to note that the outgoing photon wavefunction at the beam-splitter is a perfect superposition of basis states (as far as technical tolerances allow)
\[ \ket{\psi} = \frac{1}{\sqrt{2}}\left(\ket{1}_H + \ket{1}_V\right)\]
where $\ket{1}_H, \ket{1}_V$ denote a photon in the horizontal, vertical mode of the interferometer respectively.\footnote{Here basis states are eigenstates of photon number in each mode of the interferometer and we assume a single input photon.} A relative phase difference between returning basis states, due to a path length difference determines their complex coefficients which ultimately contributes to the squared amplitude of the state in the output mode. A photodetector, placed at the unused port which performs a measurement of the photon wavefunction detects a photon with probability $p = \sin^2(\phi/2)$, where $\phi$ is the relative phase difference.

Note how in this description, there is no noise in the interferometer which needs to be removed, nor is there any noise intrinsic to the quantum state. We do not need to worry about correlating the light intensity in each arm of the interferometer, except to manufacture an ideal beam-splitter, nor do we need to remove laser intensity fluctuations or photon shot-noise. If $N$ independent photons are inserted into the interferometer, the basis states of each individual wavefunction interfere upon return, and the summation of many individual probabilistic measurements yield the detected signal with $\sqrt{N}$ fluctuations. This noise is a statistical effect due to performing $N$ probabilistic measurements and it is this noise which is fundamentally unavoidable! As projection noise is the result of the Born rule, any proposal that surpasses this limit is in violation of quantum mechanics. From this description we can see that operating the interferometer so the unused port is dark reduces the observed laser power noise and quantum projection noise, but importantly also reduces the signal since $\frac{\partial p}{\partial \phi} \rightarrow 0$ and no sensitivity benefit is obtained \cite{Itano1993}.% Operating close to a dark fringe provides no fundamental benefit if there is no technical noise, however it can have practical advantages. If readout noise at the photodetector dominates, then operating half-way between a dark and bright fringe improves sensitivity, but if laser power noise dominates, then it is advantageous to operate close to a dark fringe. \textit{But not directly on the dark fringe as the sensitivity goes to zero. Need to check this}\footnote{What about collective or non-linear effects occurring when multiple identical photons enter at the same time? Well, these generally occur when identical photons enter at different input ports, e.g. Hong-Ou-Mandel. We don't need to worry about these because we assume nothing enters the input port.}.

\textbf{A remark:} It is a real, demonstrable effect, that one can insert light into the unused port of a beam-splitter and observe reduced laser intensity fluctuations at the interferometer output. The mechanism is destructive interference. By correctly setting the phase of this second beam, one can induce destructive interference so that less light is observed at the output. Of course under this description it is clear that such a technique will not fundamentally improve the operating sensitivity.

The above only discusses the mistaken association of quantum projection noise with vacuum fluctuation noise. How about photon shot-noise randomising the arrival time of photons, whereas the photons in squeezed light are ``\href{https://news.mit.edu/2023/ligo-surpasses-quantum-limit-1023}{holding hands}"? This is a basic error misattribution. As discussed \href{https://pubpeer.com/publications/88541E64A483E17BFF7B117D00F911}{here}, it is the relative photon distribution at the output ports rather than the photon arrival time that provides information on the phase shift.\footnote{Although a change in path length does alter the arrival time, sensitivity to this change requires both knowing the precise photon ``departure'' time and a detector with sub-femtosecond resolution, many order of magnitude below that used by LIGO.} Say I send you a message encoded in photon-wavepackets at a given communication rate. The amount of information you can extract from the received photons doesn't depend on whether I send them to you at precise, regularly spaced intervals, random intervals, or all at once at the end of the communication duration. As long as the average communication rate and the amount of information per photon is constant, then the amount of information you receive per unit time is independent of noise on the photon arrival time. Furthermore, if I reduce the randomness in message arrival time by duplicating the information into two photon-wavepackets and ensuring that they arrive equally spaced, then you actually receive \emph{less} information per unit photon (and time if power is held constant). Of course, if this duplicated wavepacket enters the communication channel through a second input port, and I don't include these photons in my accounting when defining the average communication rate, then with non-perfect detection it can appear like you receive more information per photon.\footnote{Especially if I name these extra photons squeezed vacuum. I mean, the vacuum contains no photons, so surely I'm not adding more photons to the channel and not counting them, right?}

The same argument applies if we consider information encoded in the photon arrival time. If we take a statistical distribution of photon arrival times, and substitute each photon with two photons, then all we are doing is duplicating the same information. Much better is to remove those duplicated photons and obtain new, independent statistical samples. Importantly, if a signal alters the duration it takes for each photon to reach you, then the arrival time alone does not provide any information on the signal, one also needs to know the photon departure time.

\subsection{Proof assumptions}\label{App:Assumptions}
For completeness, we list below, some assumptions of the \hyperlink{proof}{Proof} in the main text. In particular we assume:
\begin{itemize}
\item The postulates of quantum mechanics.
\item The entire dependence on $\theta$ is encoded in $\Op{H}(\theta,t)$. The starting state $\ket{\psi_0}$ and the measurement operator are independent of $\theta$.
\item Apart from the initial state-vectors, everything else in the external universe is the same when comparing unentangled and entangled ensembles. The prior information on $\theta$ and the ability to apply control fields is the same. In fact, my understanding is that the squeezing community claim even unitary evolution is the same for these two states, therefore the only difference of consequence is the starting state (and the measurement).
\item Each copy of the same state is independent and measurements on copies of the same state are identical. This assumption ensures that any enhancement over the SQL comes solely from entanglement, not other correlations, additional information or different control on copies.\\
\end{itemize}
\textbf{Comment on the first assumption:}
Using a simple example, I discuss how one assumption of the analysis presented here is critical to the proof. The assumption is the (Born rule) postulate of quantum mechanics -- the probability $\func{Pr}{x}$ of obtaining a measurement outcome $x$ when measuring a state $\ket{\psi}$ is equal to $\func{Pr}{x} = \bra{\psi}\Op{X}^{\dagger}_x \Op{X}_x\ket{\psi}$, where $\{\Op{X}_x\}$ is the collection of self-adjoint measurement operators associated with outcomes $x$ (Postulate 3 in \cite{Nielsen2000}) and $\Op{X}^{\dagger}$ denotes the Hermitian adjoint of the operator $\Op{X}$. This postulate does not account for the fact that measurements are never perfect and that some quantum states are easier to measure than others.

Take fluorescent readout of the spin state along $z$ where the state $\ket{0}$ fluoresces strongly and $\ket{1}$ does not fluoresce. If we compare readout of the following spin states:
\begin{align*}
\mathrm{State}\,1: \quad \frac{1}{\sqrt{2}}(\ket{0}& + \ket{1})\\
\mathrm{State}\,2: \;\; \frac{1}{\sqrt{2}}(\ket{00}& + \ket{11}),
\end{align*}
performed by scattering photons and counting the total number of photons detected (this is a collective spin measurement), then State\,2 is easier to measure. The photon difference between the two measurement outcomes for State\,2 is twice that when reading out State\,1.\footnote{One could argue that these measurement results have greater variance, the variance of the operator $\Op{S}_z$ i.e. $(\Delta \Op{S}_z)^2$ is greater for State\,2. Again, this operator variance argument is often incorrectly employed to say that measurement observables of one state provide more information on the spin direction than the other. This argument is incorrect at a fundamental level. If the measurements are perfect, then both observables provide the same information on the spin direction. The reason being that the definition of the measurement variance is arbitrary to some extent, it just depends on how we label the measurement outcomes. We can take readout of the first state and multiply the result by an arbitrary large number to increase the variance (or multiply by a small number to reduce the variance). In terms of measurement statistics, this does not change the information contained in the measurements. The information is independent of the labelling, just as the entropy of a probability distribution is independent of the labelling.} Whenever state readout is not perfect, then (under the fair sampling assumption) the second state provides more information. A few points of emphasis:
\begin{itemize}
\item While this is a practical limitation in every real experiment, the proposition that it represents a fundamental precision limit in quantum squeezing is equivalent to saying quantum mechanics is incorrect. 
\item Increasing the readout fidelity removes this loophole since the information difference reduces as the readout improves.
\item When people compare to the SQL and claim to surpass it, the limit is taken assuming the readout fidelity is perfect. This is what the SQL means -- the best quantum mechanics says one can possibly do, without entanglement.
\item State\,2 utilises two spins, so a fair comparison would be to compare to two independent copies of State\,1. When this comparison is performed, State\,2 provides less information.
\item In spin-squeezing literature, State\,2 (or its extension to $N$ particle Hilbert space) is not being used, the following is used instead:
\[ \frac{1}{\sqrt{2(1+\epsilon^2)}}\left(\epsilon\ket{00} + \left(\ket{10}+\ket{01}\right) + \epsilon\ket{11}\right) \]
where $\epsilon$ is real, $\epsilon <1$ for a squeezed state, $\epsilon \ll 1$ in the limit of strong squeezing, and $\ket{10}$, $\ket{01}$ cannot be distinguished by a collective spin measurement along $z$ \cite{Kitagawa1993, Ma2011, Andre2002, Vuletic2023} (see \hyperlink{FAQs}{FAQ} \ref{FAQ:1}).\\
\end{itemize}

\noindent\textbf{Assumptions for physical interpretations:}

The following assumptions are not required to derive the mathematical inequalities in the \hyperlink{proof}{Proof}, but if one then uses those inequalities to claim that the precision of a squeezed device cannot surpass the precision of an ideal unentangled ensemble, that argument requires further assumptions. I.e. when going from a purely mathematical proof to physical interpretations one makes implicit use of these assumptions. 
\begin{itemize}
\item In the proof we showed that the maximum information $\funcwrt{I}{\theta, t}{1}$ on $\theta$ provided by a single spin occurs when $|\alpha_1|^2 - |\alpha_2|^2 = 0$, which means that the initial state $\ket{\psi_0}$ of the single spin is $\ket{\psi_0} = \frac{1}{\sqrt{2}}\left(\ket{0} + e^{\iu \Phi}\ket{1}\right)$, where $\Phi$ is an arbitrary phase. In general, (quantum) measuring devices are not initialised in $\ket{\psi_0}$, therefore we should account for the time and resources required to prepare this state. Using $\ket{\psi_{0,1}}$, $\ket{\psi_{0,SS}}$ to differentiate between $\ket{\psi_0}$ for a single spin and a squeezed state respectively, then we need to assume the time and resources to generate $\ket{\psi_{0,1}}$ is not greater than $\ket{\psi_{0,SS}}$. In every proposal and experiment that I am aware of, at least one copy of $\ket{\psi_{0,1}}$ is used as an intermediary to create $\ket{\psi_{0,SS}}$. For example, to create entangled spin states, a $\pi/2$-rotation is performed on at least one spin as the first step. For Ramsey interferometry in particular, a $\pi/2$-rotation on $N$ spins creates $N$ copies of $\ket{\psi_{0,1}}$. Additional gates are then required to produce just one copy of $\ket{\psi_{0,SS}}$. To create squeezed photon states, single photons are inserted into a non-linear optical element, or in the case of LIGO, are inserted into the interferometer together with an additional field. Thus the assumption that, in general, it is more technically demanding, time-consuming and requires more resources to generate the ideal $\ket{\psi_{0,SS}}$ compared to the ideal $\ket{\psi_{0,1}}$, seems correct. In fact, when analysing the precision of entangled NOON, GHZ and CAT type states \cite{McGuinness2021}, (I claim) it precisely because the converse of this assumption does not hold that these entangled states do not outperform a single particle.\footnote{I.e. for NOON, GHZ and CAT states to reach the Heisenberg limit, the assumption is that $\ket{\psi_0}$ can be generated instantaneously with no extra resources. When the precision of these devices is presented \cite{Meyer2001, Leibfried2004, Shaniv2018, Xie2021}, the time taken to prepare $\ket{\psi_0}$ is almost always left out of the analysis, this gives a grossly distorted picture of the measurement precision.} I would like to remove this assumption, but have been unable to find a rigorous proof. One approach is to compare the distance between the initialised basis state $\ket{0}^{\otimes N}$ and $\ket{\psi_{0,1}}$ or $\ket{\psi_{0,SS}}$. However, $\func{Dist}{\ket{0},\ket{\psi_{0,1}}} \nless \func{Dist}{\ket{0}^{\otimes N},\ket{\psi_{0,SS}}}$ for any distance metric I have found.
\item The proof treats the signal parameter $\theta$ classically and assumes it has a deterministic, fixed value that does not vary. As a result, the SQL that we refer to assumes that measurement back-action onto the signal is zero. This is standard in theoretical analyses of the Heisenberg limit, especially in metrology with spins. For the proof to hold when considering measurement back-action, we need to show or assume that a single spin produces no greater measurement back-action onto the signal compared to a single squeezed state. Again it seems like this assumption can be made rigorous by proving that a single spin produces minimal back-action for the amount of information extracted, and showing this is not the case for a single squeezed state. We can remove this assumption with respect to LIGO, since a single photon has lower radiation pressure than a squeezed state, which by definition is a non-separable state of more than one photon. Experimentally, our bound corresponds to the low light power operation regime of LIGO, however we emphasise that as squeezed states produce more back-action than a single photon the results also apply to the high power regime.
\item In going from (Fisher) information to measurement uncertainty, we make assumptions on the estimator $\est{\theta}$ obtained from the measurement datum. Note, we do not need to assume that the minimal variance unbiased estimator exists, which may not always be the case \cite{Kay1993}. We just need a measurement of a single spin to provide any estimator with lower mean squared error than the best estimator obtained from readout of $\ket{\psi_{SS}(\theta,t)}$. In practice, this is easy to check, since state readout of a single spin (in a two-dimensional measurement basis) gives one of only two outcomes and finding the estimator is straightforward.
\item Whenever people claim to surpass the SQL using spin-squeezing, the limit they compare to is based on $\funcwrt{I}{\theta, t}{1}$, with a suitable multiplication by the number of independent particles. However, it is very clear to me that, due to the reasons outlined above, the information limit for an ideal measurement of a single spin ($\funcwrt{I}{\theta, t}{1}$) cannot be reached. Thus, one could modify the claim to a weaker version along the lines of ``even though squeezed states cannot surpass $\funcwrt{I}{\theta, t}{1}$, a measurement of a single spin cannot provide this much information either; if we replace by $\funcwrt{I}{\theta, t}{1}$ by a much lower limit corresponding to the precision that can be achieved \textit{in practice} with a single spin, then a measurement of a single squeezed state can surpass this limit". Note, this is not the story presented by the spin-squeezing community, you will often read about sensitivities approaching the Heisenberg limit or the conditions upon which one can saturate the limit. In order to disprove this weaker claim, one would need to prove a lower bound that is saturable. Failing that we need to assume that whatever the gap between the proven theoretical bound and what is achievable in practice, that squeezed states cannot fit in this gap. All available experimental and theoretical evidence points to this assumption being correct. In general, the finite time required to; create $\ket{\psi_0}$, rotate the state before projective readout, and perform projective readout; prevent $\funcwrt{I}{\theta,t}{1}$ from being reached. However, this is also applies to squeezed states. We need to further assume there is no other technical noise or decoherence, or more precisely that measurements of single particle states are less susceptible to this noise. Again, lower susceptibility to noise is generally observed for non-entangled states. Importantly, the assumption that $\funcwrt{I}{\theta, t}{1}$ can be reached is contained in the postulates of quantum mechanics, it allows one to define a rigorous information bound, and is generally used in derivations of the SQL and $\Delta \est{\theta}_1$. Going beyond this assumption seems to require a modification of the postulates of quantum mechanics. It is a position I strongly advocate.
\\
\end{itemize}

\subsection{Proof - Average expected information and mixed states}\label{App:Mixed}
The key to these extended proofs is that we are dealing with probability distributions over non-negative real numbers, so we don't have to worry about any complex number magic. When comparing two summations of non-negative real numbers where each term in one sum is less than each corresponding term in another sum, by convexity we have an inequality.\\

\noindent \textbf{Mixed states:} To prove the claim when the squeezed state is a mixed state $\ket{\rho_{0,SS}}$, we can write the initial state as probability distribution over $L$ pure states, with probabilities $p_l$:
\[\ket{\rho_{0,SS}} = \sum_{l=1}^L p_l \ket{\psi_{0,SS}}_l.\]
As we have assumed that the initial state does not depend on $\theta$, we can use the property that the Fisher information is convex over mixed states when $p_l$ is independent of $\theta$ \cite{Liu2020}. The fact that the convexity of the Fisher information implies that mixed states cannot increase the information bound (it is maximised by choosing pure states) has been known for a long time \cite{Mandelstam1945, Holevo1982, Braunstein1996, Giovannetti2011, Pezze2018}.\\

%Then as the information from any individual pure state does not exceed a single spin, the average information from the mixed state cannot either. This is equivalent to  Although, here we do not need to assume independence of probabilities because the optimal state independent of $p_l$.

%For mixed states, when the probabilities are parameter independent, i.e. when $\func{Pr}{\theta} = \mathrm{Pr}$, then the quantum Fisher information is convex:
%\[ \func{I}{\func{Pr}{\cdot}\rho_1(\theta)+(1-\func{Pr}{\cdot})\rho_2(\theta),H} \leq \func{Pr}{\cdot}\func{I}{\rho_1(\theta),H}+(1-\func{Pr}{\cdot})\func{I}{\rho_2(\theta),H}\]

\noindent\textbf{Average expected information:} Eq.\,\eqref{eq:FI} just provides the information around a given value of $\theta$. The expected information, averaged over the prior probability distribution of $\theta$ is:
\be \langle \func{I}{\theta,t}\rangle = \int_{\Theta} \mathrm{d}\theta \func{Pr}{\theta} \func{I}{\theta,t}, \quad \langle \func{I}{\theta,t}\rangle = \sum_{j=1}^J \func{Pr}{\theta_j} \func{I}{\theta_j,t}
\ee
where $\func{Pr}{\theta}$ is the prior probability that the parameter takes the value $\theta$. The LHS expression is the expected information when the prior probability is continuous, taking on values in $\Theta$, and the RHS when $\theta$ can take $J$ discrete values.

Given that we have proved for any value of $\theta$, the information $\funcwrt{I}{\theta, t}{SS}$ is not greater than that for a single spin $\funcwrt{I}{\theta, t}{1}$, then we can also use convexity to show that this holds for the expected information averaged over the prior probability distribution of $\theta$.\\% Another method is to note that the optimal initial state is independent of the value of $\theta$ and therefore also independent of the prior probability of $\theta$. It may not be the case that the optimal measurement is independent of prior probability distribution of $\theta$. To additionally prove that no measurement can produce more information, we use the result of Giovannetti, Lloyd and Maccone who showed that entanglement does not provide any advantage at the measurement stage \cite{Giovannetti2006}. Thus we only need to consider the amount of information encoded in the state (before measurement).\\

\subsection{Proof - Arbitrary time-dependent Hamiltonian: $\mathbf{H}(\theta, t)$}\label{App:General}
Let $\Op{H}^1(\theta,t)$ be an arbitrary time-dependent Hamiltonian acting on a single spin-$1/2$ particle, i.e. 2-dimensional Hilbert space. Assume the spectrum of $\Op{H}^1(\theta,t)$ is bounded. Then any spin-1/2 state $\ket{\psi_1}$ has \hyperlink{QFI}{quantum Fisher information} on $\theta$ bounded by
\be \func{QFI}{\ket{\psi_1},\theta} \leq \left(t \cdot \mu_{\delta}/\hbar\right)^2 \label{eq:QFIsin} \ee
where $\mu_\delta$ is the difference in eigenvalues of $\ket{\psi_1}$ with respect to the operator $\frac{\partial \Op{H}^1(\theta,t)}{\partial \theta}$. \Cref{eq:QFIsin} is saturated by placing the spin in an equal superposition of eigenstates of $\frac{\partial \Op{H}^1(\theta,t)}{\partial \theta}$. Let $E_1$ denote the maximal energy difference of the state with maximal quantum Fisher information. Now let $\Op{H}^N(\theta,t)$ denote the Hamiltonian interacting with $N$ identical spin-1/2 particles. The QFI bound is saturated by placing the $N$ particles in a single non-separable state with $N$ times greater value of $\mu_{\delta}$. The energy gap of this state is also $N$ times greater than $E_1$. Any state with energy gap bounded by $E_1$ obeys the same bound on QFI as the spin-$1/2$ particle.

More generally, let $\delta\ket{\psi}/\delta\theta$ denote the change in Hilbert space angle of $\ket{\psi}$ for an infinitesimal change in $\theta$. If this quantity is the same for two non-separable quantum states, then so is the quantum Fisher information on $\theta$.\\% \textcolor{blue}{This quantity is the quantum Fisher information of $\ket{\psi}$ on $\theta$.}\\

For the main text, we motivate the proof by noting that if we could explicitly solve the unitary evolution of a single spin and show that it remains in the optimal state for all evolution times, then this would imply $\left|\bra{\psi_1(\theta,t)}\left(\frac{\mathrm{d}\ket{\psi_1(\theta,t)}}{\mathrm{d} \theta}\right)\right|^2 = 0$. However, without a general method of solving time-evolution for arbitrary Hamiltonians, even when restricting ourselves to a single spin, we cannot make that statement. An equivalent approach is to use the information bound proved by Pang and Jordan for general time-dependent Hamiltonians \cite{Pang2017}:
\be
\func{I}{\theta,t} \leq \left| \int^t_{0} \mathrm{d}t' \left(\mu_{\mathrm{max}}(t') - \mu_{\mathrm{min}}(t')\right)/\hbar \,\right|^{2},
\label{eq:Pang}
\ee
where $\mu_{\mathrm{max}}(t), \; \mu_{\mathrm{min}}(t)$ are the maximum, minimum eigenvalues of $\frac{\partial \Op{H}(\theta,t)}{\partial \theta}$. Eq.\,\eqref{eq:Pang} expresses a simple concept, the information per unit time is limited by the maximum response speed of the meter to changes in $\theta$. This is quantified by the eigenvalues of $\frac{\partial \Op{H}(\theta,t)}{\partial \theta}$ (c.f. how the eigenvalues of $\Op{H}$ determine the evolution speed of any quantum state). Minimising the uncertainty requires placing the meter in an equal superposition of states with extremal eigenvalues, $\mu_{\mathrm{max}}(t), \; \mu_{\mathrm{min}}(t)$:
\[
\ket{\psi^{\mathrm{Opt}}(t)} = \frac{1}{\sqrt{2}}\left(\ket{\mu_{\mathrm{max}}(t)} + e^{\iu \Phi}\ket{\mu_{\mathrm{min}}(t)}\right),
\]
where $\Phi$ is an arbitrary phase \cite{Braunstein1996, Giovannetti2006, Childs2000}.

%\textbf{Already this analysis is sufficient to prove the claim...} But the argument is not very mathematical

If we could solve the unitary evolution under $\Op{H}(\theta, t)$, show that $\ket{\psi_1(\theta,t)} = \ket{\psi^{\mathrm{Opt}}(t)}$ for all time, and find a projective measurement that saturates the bound, then we can saturate Eq.\,\eqref{eq:Pang} and show that $\funcwrt{I}{\theta,t}{1} = 4\left(\frac{\mathrm{d}\bra{\psi_1(\theta,t)}}{\mathrm{d} \theta}\right)\left(\frac{\mathrm{d}\ket{\psi_1(\theta,t)}}{\mathrm{d} \theta}\right)$. Unfortunately in general this is not possible: for example time-dependent Hamiltonians where both the eigenvalues and eigenstates of $\Op{H}$ depend on $\theta$. Therefore we need to modify our claim to address not the gradient of $\ket{\psi}$, but rather the projection of $\frac{\mathrm{d}\ket{\psi}}{\mathrm{d} \theta}$ orthogonal to $\ket{\psi}$: $\left(\frac{\mathrm{d}\ket{\psi}}{\mathrm{d} \theta}\right)_{\perp} \equiv \frac{\mathrm{d}\ket{\psi}}{\mathrm{d}\theta}-\ket{\psi}\bra{\psi}\left(\frac{\mathrm{d}\ket{\psi}}{\mathrm{d}\theta}\right)$.\\

\textbf{Claim:} If $\left(\frac{\mathrm{d}\bra{\psi_{SS}(\theta,t)}}{\mathrm{d} \theta}\right)_{\perp}\left(\frac{\mathrm{d}\ket{\psi_{SS}(\theta,t)}}{\mathrm{d} \theta}\right)_{\perp} = \left(\frac{\mathrm{d}\bra{\psi_1(\theta,t)}}{\mathrm{d} \theta}\right)_{\perp}\left(\frac{\mathrm{d}\ket{\psi_1(\theta,t)}}{\mathrm{d} \theta}\right)_{\perp}$ then $\funcwrt{I}{\theta,t}{SS} = \funcwrt{I}{\theta,t}{1}$.\\

\textbf{Proof:} As noted by Braunstein, Caves and Milburn \cite{Braunstein1996}
\[ \left(\frac{\mathrm{d}\bra{\psi(\theta,t)}}{\mathrm{d} \theta}\right)_{\perp}\left(\frac{\mathrm{d}\ket{\psi(\theta,t)}}{\mathrm{d} \theta}\right)_{\perp} = \left(\frac{\mathrm{d}\bra{\psi(\theta,t)}}{\mathrm{d} \theta}\right)\left(\frac{\mathrm{d}\ket{\psi(\theta,t)}}{\mathrm{d} \theta}\right) - \left|\bra{\psi(\theta,t)}\left(\frac{\mathrm{d}\ket{\psi(\theta,t)}}{\mathrm{d} \theta}\right)\right|^2 \]
and Eq.\,\eqref{eq:BCM} proves the claim. This statement is equivalent to looking at the data-set $\{x_i;\theta,t\}$ resulting from projective measurements of $\ket{\psi_{SS}(\theta,t)}$. If this data-set shows no greater dependence on $\theta$ than a single spin, then it cannot contain fundamentally more information on $\theta$ than projective measurements of a single spin. Again, this is basic statistical analysis, it is simply a restatement of classical Cram\'er-Rao bound, for a probabilistic data-set.

\subsection{Glossary}\label{Glossary}
\begin{definition}[\hypertarget{non-separable}{Non-separable} quantum state]
A quantum state $\ket{\psi}$, a unit length vector in a complex Hilbert space $\mathcal{H}$, is defined as a \emph{non-separable} quantum state if and only if it cannot be decomposed into the (outer) tensor product of two quantum states, each in Hilbert spaces of dimension greater than 1 (and lower dimension than $\mathcal{H}$). Conversely, if
\[ \ket{\psi} = \ket{\psi_1} \otimes \ket{\psi_2}, \text{ where } \ket{\psi_1} \in \mathcal{H}_1 \text{ with } \func{dim}{\mathcal{H}_1} > 1, \;  \text{ and } \ket{\psi_2} \in \mathcal{H}_2 \text{ with } \func{dim}{\mathcal{H}_2} > 1,\]
then $\ket{\psi}$ is a \emph{separable} quantum state.
\end{definition}
More fully, since the two vectors $\ket{\psi}$ and $e^{\iu \theta}\ket{\psi}$ differing by a global phase are thought to be indistinguishable, we consider quantum states as unit length `rays'. That is why the quantum Fisher information in \cref{eq:QFI} considers the energy \emph{difference} between eigenstates and the form given in \cref{eq:BCM} includes $\left|\bra{\psi(\theta,t)}\left(\frac{\mathrm{d}\ket{\psi(\theta,t)}}{\mathrm{d} \theta}\right)\right|^2$ so that multiplication of a state with a global phase dependent on $\theta$ does not increase the QFI of that state.

\begin{remark}[Non-separable states]
There is a clear connection between non-separable quantum states and prime numbers -- they both cannot be factored into smaller units. This suggests that we treat non-separable quantum states as the atomic or indivisible units in our model. Note also that just as the trivial decomposition of a prime number by a factor of 1 does not make the number composite, the tensor product of a non-separable quantum state with a unit dimension Hilbert space does not produce a separable state vector. For example, in optical interferometry quantum states are often written in the occupation number basis $\ket{n}$, with $n$ denoting the number of particles in a given mode. Sometimes the empty or vacuum state of another mode is considered and such a state would be written $\ket{n}\otimes\ket{0}$. If we only consider the vacuum state and never add any photons to this mode, then the Hilbert space of this mode has unit dimension and can be disregarded allowing us to write $\ket{n}$ without loss of generality.
\end{remark}

This exclusion of Hilbert spaces with unit dimension does away with the issues raised by Seth Lloyd that entanglement is not intrinsic to a state, but rather is a property of how it is partitioned \cite{Lloyd1999}.\footnote{I thank discussions with Warwick Bowen for bringing this to my attention.} In particular, here we treat a single photon in a superposition of two modes
\[ \ket{\psi} := \frac{1}{\sqrt{2}}\left(\ket{1}_V + \ket{1}_H\right) \]
as non-entangled state, even though it can also be written as
\[ \ket{\psi} := \frac{1}{\sqrt{2}}\left(\ket{1}_V\ket{0}_H + \ket{0}_V\ket{1}_H\right), \]
sometimes called a path-entangled state.

\begin{definition}[Classical \hypertarget{Fisher information}{Fisher information}]
Let $x$ be a single measurement datum sampled from probability distribution $\func{Pr}{x;\theta}$. The (classical) Fisher information on $\theta$ is defined as:%, the derivative is calculated at the true value of $\theta$. The equality holds assuming that we use an unbiased estimator, the measurement operator $\Op{X}$ corresponding to the observable $x$ does not depend on $\theta$, and we can interchange the trace and the derivative to write $\frac{\partial}{\partial\theta}\func{Tr}{\rho \Op{X}} = \func{Tr}{\frac{\partial \rho}{\partial \theta} \Op{X}}$.}:
\[ \func{I}{\theta} := \8{\left(\frac{\partial \ln\left[\func{Pr}{x;\theta}\right]}{\partial \theta}\right)^2}.\]
This can be shown equal to 
\[
\func{I}{\theta} := \int_{X} \mathrm{d}x \frac{1}{\func{Pr}{x;\theta}}\left(\frac{\partial \func{Pr}{x;\theta}}{\partial \theta}\right)^2, \quad \func{I}{\theta} := \sum_{i=1}^R \frac{1}{\func{Pr}{x_i;\theta}}\left(\frac{\partial \func{Pr}{x_i;\theta}}{\partial \theta}\right)^2.
\]
for continuous and discrete probability distributions.
\end{definition}

\begin{definition}[\hypertarget{QFI}{Quantum Fisher information}]
For a pure non-separable quantum state $\ket{\psi(\theta,t)}$ with dependence on $\theta$, the \emph{quantum Fisher information} on $\theta$ is defined as
\[\func{QFI}{\theta,t} := 4\left[\left(\frac{\mathrm{d}\bra{\psi(\theta,t)}}{\mathrm{d} \theta}\right)\left(\frac{\mathrm{d}\ket{\psi(\theta,t)}}{\mathrm{d} \theta}\right) - \left|\bra{\psi(\theta,t)}\left(\frac{\mathrm{d}\ket{\psi(\theta,t)}}{\mathrm{d} \theta}\right)\right|^2\right].\]
This quantity bound the classical Fisher information of any measurement on $\ket{\psi(\theta,t)}$.
\end{definition}

\section*{(In)Frequently Asked Questions}\hypertarget{FAQs}{}
\setcounter{subsection}{0}
\subsection{I find it difficult to accept your statement that the whole field is wrong without more evidence. Could you help point out some specific errors in papers.}\label{FAQ:1}

I have started doing that for many experiments, see Refs.\,\cite{McGuinness2021, McGuinness2022, McGuinness2023, McGuinness2023a}. Let's go through an explicit example for spin-squeezing. All you need to do is ask some questions about what is going on and see if the answers make sense. Some questions that might help:
\begin{quotation}
In spin-squeezing, what entangled state is being created?
\end{quotation}
It is surprisingly difficult to find a direct answer to this question. Have a look through these papers and see if you can find an explicit description of the squeezed state anywhere \cite{Orzel2001, Esteve2008, Appel2009, Eberle2010, Gross2010, Koschorreck2010, Koschorreck2010a, Leroux2010, Leroux2010a, LouchetChauvet2010, Riedel2010, SchleierSmith2010, Wasilewski2010, Luecke2011, Sewell2012, Bohnet2014, Muessel2014, Strobel2014, Cox2016, Davis2016, Hosten2016, Hosten2016a, Kruse2016, Linnemann2016, Macri2016, Braverman2019, Bao2020, Malia2020, PedrozoPenafiel2020, Schulte2020, Colombo2022, Greve2022, Malia2022, Bornet2023, Eckner2023, Franke2023}. Even the review articles on spin-squeezing \cite{Gross2012, Pezze2018} do not give an example of a squeezed state-vector.\footnote{You will find a definition of a coherent spin state as the tensor product of $N$ qubits in pure states pointing along the same mean spin direction, but no example of any state that could qualify as a squeezed state. Doesn't it seem strange that none of these papers define the quantum state they are using?} %Abadie2011, Asai2013, Tse2019,
In place of a mathematical definition, a pictorial representation of the squeezed state on the Bloch sphere is used. I propose a simple explanation as to why the spin-squeezed state is rarely defined. No state exists that fulfils the two key properties claimed by the spin-squeezing community; \textit{i)} reduced measurement uncertainty, and \textit{ii)} no enhanced state response to the signal. Whenever such a candidate state is presented, it is easy to confirm that it does not have the required properties.

For a system consisting of two particles, Mark Kasevich and Vladan Vuleti\'c inform me that the state that everyone is preparing is \cite{Vuletic2023}:
\be \frac{1}{\sqrt{2(1+\epsilon^2)}}\left(\epsilon\textcolor{green!50!black}{\ket{00}} + \left(\textcolor{red!50!black}{\ket{10}+\ket{01}}\right) + \epsilon\textcolor{green!50!black}{\ket{11}}\right)
\label{eq:squeezed}
\ee
where $\epsilon$ is real, $\epsilon <1$ for a squeezed state, $\epsilon \ll 1$ in the limit of strong squeezing, and $\ket{10}$, $\ket{01}$ cannot be distinguished by a collective spin measurement along $z$ (I have added colours for emphasis). Equivalent definitions and the extension to $N$ particles are given in \cite{Ma2011, Kitagawa1993,Andre2002}. If the state is highly squeezed ($\epsilon \approx 0$) we can approximate it as
\[\ket{\psi_{0,SS}} = \frac{1}{\sqrt{2}}\left(\ket{10}+\ket{01}\right).\]
Let's assume $\ket{\psi_{0,SS}}$ is the input state to a Ramsey interferometer, where a relative phase $\varphi$ accumulates between the $\ket{10}$ and $\ket{01}$ basis states.\footnote{Forget about what type of Hamiltonian is required to generate such a phase response and whether the experiments actually implement such a Hamiltonian.} Then, $\ket{\psi_{0,SS}}$ evolves to $\ket{\psi_{SS}(t)} = \frac{1}{\sqrt{2}}\left(\ket{10}+e^{\iu \varphi} \ket{01}\right)$. The quantum phase $\varphi$ is compared to the phase of the final $\pi/2$-pulse in the interferometer, and assuming that the $\pi$/2-pulses are strong and have the same phase in the rotating frame, the output state of the Ramsey interferometer is approximately
\[\ket{\psi_{SS}(t)} \approx \frac{1}{2}\left[\left(1-e^{\iu\varphi}\right)\ket{10}+\left(1+e^{\iu\varphi}\right)\ket{01}\right].\]
Now what measurement can be used to estimate $\varphi$?

In Refs.\,\cite{Appel2009, Gross2010, Koschorreck2010, Koschorreck2010a, Leroux2010, Leroux2010a, LouchetChauvet2010, Riedel2010, SchleierSmith2010, Luecke2011, Sewell2012, Bohnet2014, Muessel2014, Strobel2014, Cox2016, Davis2016, Hosten2016, Hosten2016a, Linnemann2016, Macri2016, Braverman2019, Malia2020, Bao2020, PedrozoPenafiel2020, Greve2022, Malia2022, Colombo2022, Bornet2023, Eckner2023, Franke2023}, a collective spin measurement is performed. This is a projective measurement on all of the spins, followed by summing up the measurement results. I.e. just count the number of spins measured in the up state. A collective spin measurement along $z$ is
\[\tilde{\Op{S}}_z = \sum_{i=1}^N \Op{S}_z,\]
where we will label the individual measurement outcomes of $\Op{S}_z$ as ${S}_z \in \{\pm \frac{1}{2}\}$. You can confirm for yourself, that this measurement on $\ket{\psi_{SS}(t)}$ always returns the same result: $\tilde{\Op{S}}_z = 0$ independent of the phase. The most important point to highlight about this measurement is that it gives no information on $\varphi$. This measurement of $\ket{\psi_{SS}(t)}$ is metrologically useless, because one always obtains the same measurement result $\tilde{S}_z = 0$ regardless of any phase shifts.

In this example it is easy to see how one can claim to make a state where the variance of $\tilde{\Op{S}}_z$ (or the variance of the data-set $\tilde{S}_z$) reduces or goes to zero, while the perpendicular spin vector length is not reduced (it points in the $x-y$ plane). Of course the variance goes to zero, because the measurement always returns the same result. This is precisely the claim made in the following experimental works \cite{Esteve2008, Appel2009, Gross2010, Leroux2010, Leroux2010a, LouchetChauvet2010, Riedel2010, SchleierSmith2010, Wasilewski2010, Luecke2011, Sewell2012, Bohnet2014, Muessel2014, Strobel2014, Cox2016, Davis2016, Hosten2016, Hosten2016a, Macri2016, Braverman2019, Bao2020, Malia2020, PedrozoPenafiel2020, Colombo2022, Greve2022, Malia2022, Bornet2023, Eckner2023, Franke2023}.\footnote{These claims are based on the definition of squeezing given in Ref.\,\cite{Wineland1992}, and often related to the Wineland squeezing parameter.} This analysis highlights a basic error in the analysis of the squeezing community, clearly $\Delta \varphi \neq \sqrt{\func{Var}{\tilde{\Op{S}}_z}}$. The uncertainty in the phase is not reduced, because the measurement observable does not depend on the phase. In these works \cite{Esteve2008, Appel2009, Gross2010, Leroux2010, Leroux2010a, LouchetChauvet2010, Riedel2010, SchleierSmith2010, Wasilewski2010, Luecke2011, Sewell2012, Bohnet2014, Muessel2014, Strobel2014, Cox2016, Davis2016, Hosten2016, Hosten2016a, Macri2016, Braverman2019, Bao2020, Malia2020, PedrozoPenafiel2020, Colombo2022, Greve2022, Malia2022, Bornet2023, Eckner2023, Franke2023}, low variance measurements of $\tilde{\Op{S}}_z$ form the primary piece of evidence that $\varphi$ has been estimated with an enhanced precision.

%Even when the squeezing is not perfect, there are several properties of the data from collective spin measurements that mean an improved estimate of $\varphi$ is not possible.
%\begin{itemize}
%\item For $\epsilon = 1$, although one might say the state in Eq.\,\eqref{eq:squeezed} is not squeezed, it is still entangled. On average, we would expect to observe Ramsey fringes with a different contrast and period compared to the same number of independent spins. For a two-spin ensemble, if the unentangled ensemble has a period of $2 \pi$ and the fringe contrast goes from $1$ to $-1$, then the entangled ensemble should have a period of $\pi$  and fringes going from $1/2$ to $-1/2$. Again, this is not observed in the data.
%\item When squeezing is increased, i.e. $\epsilon \rightarrow 0$, we would expect to see contrast of the Ramsey fringes disappear when performing a collective spin measurement. It is true that the measurement variance is reduced, since we always obtain the same result $\tilde{S}_z = 0$, but this gives no information on the phase of the final readout pulse. Think about what is happening here, squeezing just reduces the amplitude of the Ramsey fringes, without affecting the period. How can this improve the measurement precision?
%\end{itemize}

Note, the variance does not need to reduce for all values of $\varphi$. In \cite{Kruse2016, Linnemann2016, Braverman2019, PedrozoPenafiel2020, Eckner2023} an initial weak measurement is performed so that when a second state measured is performed the outcomes depend on $\varphi$. The measurement basis is chosen so that $\ket{\psi_{SS}(t)}$ is an eigenstate of the measurement operator for particular values of $\varphi$, and the variance reduces just at these values (e.g. when $\varphi = 0, \pi$). What the authors do not mention is that precisely the same is observed for measurements of a non-entangled state (see \cite{Itano1993}, Fig.\,11) and that the sensitivity to $\varphi$ actually reduces at these points (this point is often obscured by analytic errors).

Finally, one might argue, that even if these measurement statistics do not allow for an improved estimation of $\varphi$, they are an ``entanglement witness" \cite{Soerensen2001, Bohnet2016, Cox2016, Eckner2023, Franke2023} in that they cannot be replicated with non-entangled states. Thus they correspond to a kind of Bell inequality. This is absolutely incorrect. The key point to note here is that these low variance measurements take information on $\varphi$, contained in some quantum state and destroys it. There are actually plenty of measurements on non-entangled states that allow one to throw away information and reproduce these measurement statistics. Take a measurement of the null Hermitian operator, which just maps any quantum state to the observable `0' with unit probability. This is clearly an allowable measurement and it also produces measurement statistics that violate the standard quantum limit as per the definition commonly used in the spin-squeezing community. Another allowable measurement is: perform a $\Op{S}_z$ measurement, flip the spin state with a $\pi$-pulse, perform a second $\Op{S}_z$ measurement, and then add these two measurement outcomes. Again, this summation is efficient at destroying any information contained in the state and would qualify as a measurement of an entangled squeezed state according to many classifications; indeed that measurement is done in \cite{Hosten2016, Malia2020, Malia2022}. Finally, a collective spin measurement is a form of post-processing that throws away extra information contained in individual measurement statistics. This process is most clearly seen in experiments where it is possible to readout the spin-states individually, but these individual data points are summed together, as in \cite{Bornet2023}.\\% These comments do not only apply to measurements on squeezed states in quantum metrology, in fact, the same type of measurements can be performed to demonstrate violation of a Bell inequality without entanglement. In particular, because the Bell inequalities are based purely on measurement variances (and not the information contained in the measurements) they can be violated by these types of measurements. When formulated in terms of information inequalities it is possible to prove and demonstrate a separation between entangled and non-entangled particles. The information inequalities, showing that one system contains more information than the other are robust to the loophole described here because, while we are always free to throw away information, we cannot create more. Assuming data processing is allowed, there is no equivalent restriction on measurement variances. The asymmetry in information processing does not apply to variances because there is always the possibility to either increase or decrease the measurement variance by an arbitrary amount. However, it is important to note that, when expressed as an information inequality, it is the system of non-entangled particles that contain more information.\\

\subsection{What squeezed-state is being prepared? What experimental evidence is presented, to verify the creation of this state?}\label{FAQ:2}

Let's assume that the state in Eq.\,\eqref{eq:squeezed} or its extension to the $N$ particle spin-squeezed state is prepared, and following \cite{Appel2009, Gross2010, Koschorreck2010, Koschorreck2010a, Leroux2010, Leroux2010a, LouchetChauvet2010, Riedel2010, SchleierSmith2010, Luecke2011, Sewell2012, Bohnet2014, Muessel2014, Strobel2014, Cox2016, Davis2016, Hosten2016, Hosten2016a, Linnemann2016, Macri2016, Braverman2019, Malia2020, Bao2020, PedrozoPenafiel2020, Greve2022, Malia2022, Colombo2022, Bornet2023, Eckner2023, Franke2023}, a collective spin measurement is performed. It is important to highlight that the measurement data presented in these papers do not agree with what one predicts for measurements on this state. In particular, there is one important characteristic of this state which is in complete discrepancy with the data presented in the literature.
\begin{itemize}
\item The green basis states accumulate a greater relative phase than that of a single particle. If the $N$ particle squeezed state has components: $\textcolor{green!50!black}{\ket{0}^{\otimes N}}$ and $\textcolor{green!50!black}{\ket{1}^{\otimes N}}$, they accumulate a relative phase, $N$ times that of a single particle. This is never observed in the data!
\end{itemize}
Note, the relative phase accumulated by the red basis states is not observable by a collective spin measurement along $z$. In fact, a perfect measurement of these basis states always gives the same result $\tilde{S}_z = 0$, independent of the phase of the final readout pulse. So for arbitrary squeezing amounts, the only observable phase dependence comes from the green basis states, whereas the red basis states -- which squeezed states apparently use as a resource -- only reduce the Ramsey fringe contrast.

Again, for emphasis. Measurements of this state in Eq.\,\eqref{eq:squeezed} give completely different data to what is presented by experiments claiming to produce spin squeezed states. In particular, if the squeezed state Eq.\,\eqref{eq:squeezed} is produced, then the \textit{period} of the Ramsey fringes should be reduced compared to the unentangled state, as well as the contrast. Strikingly, when a plot of the measurement observable vs. readout phase is presented, the period is the same as for the unentangled ensemble, see:
\begin{enumerate}
\item Figure\,2b of \cite{Gross2010}.
\item Figure\,2 of \cite{Leroux2010}. Data is only presented for the variance as a function of rotation angle. The key observation to make is that the period of the variance does not change.
\item Figure\,5 of \cite{LouchetChauvet2010}.
\item Supplementary Figure\,A6 of \cite{SchleierSmith2010}.
\item Figure\,2 and Figure\,3c of \cite{Bohnet2014}. In Figure\,2 the phase shift for a non-entangled ensemble (on the left) is compared to that of a squeezed ensemble (on the right). If anything, the non-entangled ensemble experiences a greater phase shift.
\item Figure\,1c and Figure\,2c of \cite{Muessel2014}. No data is presented for the unsqueezed ensemble, but the expected dependence on the readout phase is the same as presented for the squeezed ensemble (a period of $2\pi$).
\item Figure\,4A of \cite{Strobel2014}. This data corresponds to measurements on a state, which is prepared as a spin-squeezed state and evolved for 25\,ms. After 25\,ms squeezing is lost, although entanglement remains?! In Figure\,2, data is presented for squeezing as a function of tomography angle. 
\item Figure\,3(d) of \cite{Cox2016}. A plot of spin noise as a function of the phase $\psi$ of the final readout pulse. The phase dependence is the same as one would obtain with a non-entangled ensemble.
\item Figure\,2 of \cite{Linnemann2016} where a $2\pi$ phase dependence is observed, exactly as one expects for a non-entangled ensemble. See also Figure\,3(a) where the inset covers the full $0 - 2\pi$ range.
\item Figure\,2 of \cite{Braverman2019}, data is presented for the variance as a function of rotation angle. The dependence of the variance on the rotation angle is what one expects for a non-entangled ensemble, and the period does not change with squeezing.
\item Figure\,2c of \cite{PedrozoPenafiel2020}, a plot of variance as a function of tomography angle is presented for the spin-squeezed state. This is exactly the same angular dependence one expects for a non-entangled ensemble, in particular the period of the plot.
\item Figure\,2a of \cite{Riedel2010}, the same angular dependence is observed for squeezed and unsqueezed ensembles.
\item Figure\,4c of \cite{Greve2022} where the phase response for a squeezed and unsqueezed ensemble are directly compared. See also Figure\,4b for a plot of spectroscopic enhancement as a function of rotation angle for the spin-squeezed state. Again this phase dependence is precisely what one would expect for a non-entangled ensemble.
\item Figure\,2a (inset) of \cite{Eckner2023} where the phase response for a squeezed and unsqueezed ensemble are superposed.
\item Figure\,4a of \cite{Franke2023} where the phase response for a squeezed and unsqueezed ensemble are plotted.
\end{enumerate}
Note for measurements of noise or variance, twice the angular or phase dependence is observed compared to population measurements (i.e. half the period). The same is also observed for measurements of unentangled ensembles when the correct rotation direction is chosen. Sometimes the sensor response is pictorially represented, and no experimental data is presented, see: 
\begin{enumerate}
\setcounter{enumi}{15}
\item Figure\,1 of \cite{Appel2009}.
\item Figure\,2 of \cite{Leroux2010a}.\\
\end{enumerate}
When ``squeezing" is increased in these papers, although the measurement variance reduces it is clear that this leads to \textit{worse} phase estimation, because it simply results from contrast reduction. No evidence of a state with the characteristics predicted by Eq.\,\eqref{eq:squeezed} (except that the Ramsey fringe contrast and the measurement variance reduces) is presented.

In fact, these two observations of the spin-squeezing literature: a general lack of a mathematical definition of the squeezed state, and the same spin response as an unentangled state, are what forced me to take the approach I did. If I had instead been given the definition of a spin-squeezed state as in Eq.\,\eqref{eq:squeezed} then my approach would have been to address the fact that we expect this state to display a (partially) greater response to the signal i.e. shorter period, despite the spin-squeezing papers claiming the contrary and providing contradictory data. But I suspect the obvious contradiction between the definition in Eq.\,\eqref{eq:squeezed} and the experimental data explains why such a description of squeezed states is not usually presented. What should be abundantly clear, is that the squeezed state defined in Eq.\,\eqref{eq:squeezed}, or the $N$ particle equivalent is not created in any of the cited spin squeezing papers. So, the natural question to ask is
\begin{quotation}
What state is being created?
\end{quotation}
Summarizing the experimental evidence. Collective spin measurements of ``squeezed" ensembles have the same phase dependence as unentangled ensembles with lower contrast. To me this sounds remarkably similar to preparation of an unentangled ensemble with poor fidelity and I think that this should be the default description unless further evidence is provided. We can also read the description of how the ``squeezed" state is created. In the following papers, a ``squeezed" state is prepared by performing a QND measurement on the ensemble, i.e. a projective measurement of the spin population along $z$ \cite{Appel2009, Leroux2010, Leroux2010a, LouchetChauvet2010, SchleierSmith2010, Bohnet2014, Cox2016, Hosten2016, Braverman2019, Malia2020, Bao2020, PedrozoPenafiel2020, Greve2022, Malia2022, Hosten2016a, Colombo2022}. In general, this measurement is weak, in that it only projects a small proportion of spins to the $\ket{0}, \ket{1}$ eigenstates of $z$. After performing the QND, the authors state that the ensemble is now in an entangled ``squeezed" state.

Based on the evidence presented in these papers, this is definitely not the description I would use. I would describe some of the spins as being in $\ket{0}$, and some in $\ket{1}$, conditional on the QND measurement result. That is basic quantum mechanics, after a projective measurement, the system is in an eigenstate of the measurement observable. The spins that are not measured, are unaffected except to experience an a.c. Stark shift from the QND measurement. So, this is my explanation of what is happening in the above papers. The spins are prepared in an unentangled state: $\ket{x}^{\otimes N}$. Some of these spins are projected to $\ket{0}$, $\ket{1}$. The resulting (probability mixture) ensemble can be described as one where most of the spins are initialised to $\ket{x}$ and some are in $\ket{0}$, $\ket{1}$. This ensemble is then used to perform Ramsey interferometry. The resulting contrast is worse than an ensemble initialised to $\ket{x}^{\otimes N}$ and the phase response is the same. As far as I am aware, this is the only theoretically consistent description that fits to all of the evidence presented. Importantly this description does not involve entanglement.\\

\subsection{Don't optical two-mode squeezed states have greater quantum Fisher information?}
A counter-argument that I have encountered is the presentation of certain squeezed states of light, where it can be shown mathematically that these states have increased quantum Fisher information (for sensing a phase shift in an interferometer).\footnote{I am indebted to discussions with Leon Bello and Sultan Abdul Wadood for pointing this out and providing definitions of single-mode and two-mode squeezed vacuum states.} One of these states is the two-mode squeezed vacuum state.%, and mathematically one can show that the quantum Fisher information of this state on a phase shift in the interferometer is greater than a single photon state.

\begin{definition}[Single-mode and two-mode squeezed vacuum state]
A \emph{single-mode squeezed vacuum state} $\ket{SMSV}$ is defined as
\be
\ket{SMSV} = \frac{1}{\sqrt{\func{cosh}{r}}}\sum^{\infty}_{n = 0} \left(-e^{\iu \theta} \func{tanh}{r}\right)^n \frac{\sqrt{(2n)!}}{2^n n!}\ket{2n},
\ee
where $\ket{2n}$ denotes a state containing $2n$ photons in a single optical mode and $r$ is often referred to as the squeezing parameter. A \emph{two-mode squeezed vacuum state} $\ket{TMSV}$ is defined as
\[ \ket{TMSV} = \frac{1}{\func{cosh}{r}} \sum_{n=0}^\infty (-e^{i\theta} \tanh r)^n \ket{n,n}, \]
where $\ket{n,n}$ denotes two optical modes (often the vertical and horizontal modes of a Michelson interferometer) each containing $n$ photons. Each photon is assumed to have energy $E_1 = \hbar\omega$ where $\omega$ is the optical angular frequency. 
\end{definition}

Writing the Hamiltonian for any state that evolves through the interferometer as
\[ \Op{H}^{\text{MI}}(\phi) = \phi\cdot \begin{pmatrix} \Op{n} E_1 & 0 \\
0 & 0
\end{pmatrix} = \phi\cdot \begin{pmatrix} \Op{n} \hbar \omega & 0 \\
0 & 0
\end{pmatrix},\]
where $E_1 = \hbar \omega$ is the energy of a single photon state with frequency $\omega$, and the operator $\Op{n} = \Op{a}^\dagger\Op{a}$ denotes the $n$ photon state or a single photon with $n$ times greater energy. This Hamiltonian describes two modes corresponding to the two arms of an interferometer, where one mode experiences a phase shift $\phi$ (the parameter in optical phase shift detection is often denoted $\phi$).

The two basis states of this Hamiltonian are $\ket{V}, \ket{H}$
\[ \begin{pmatrix}
1 \\
0
\end{pmatrix} \equiv \ket{V}, \quad
\begin{pmatrix}
0 \\
1
\end{pmatrix} \equiv \ket{H},
\]
and the notation $\ket{n,m}$ is often used to represent an $n$ photon quantum state in the vertical mode and $m$ photon state in the horizontal mode, equivalently one photon of $n$ (resp. $m$) times higher energy. % basis states containing $n$ photons are denoted $\ket{n}_V, \ket{n}_H$.% With $\Op{n}$ which correspond to an interferometer containing just a single photon of energy $E_1$ in each mode.% For basis states with higher energy, we use the notation $\ket{n}_V$ to denote an eigenstate of $\Op{H}^{\text{MI}}(\phi)$ with $n$ photons or $n$ times greater energy in the vertical mode, and $\ket{n}_H$ to denote an eigenstate of $\Op{H}^{\text{MI}}(\phi)$ with $n$ times greater energy in the horizontal mode.
Under $\Op{H}^{\text{MI}}(\phi)$ the single photon state $\ket{\psi}_1 = \frac{1}{\sqrt{2}}\left(\ket{1}_H+\ket{1}_V\right)$ evolves to $\ket{\psi(\phi,t)}_1 = \frac{1}{\sqrt{2}}\left(\ket{1}_H+e^{-\iu \phi \omega t}\ket{1}_V \right)$ and the quantum Fisher information of this state is
\[ \func{QFI}{\ket{\psi(\phi,t)}_1} = \left(t \omega\right)^2.\]

When inserting the single-mode squeezed vacuum state into \emph{just the vertical mode} of the interferometer, the quantum Fisher information of this state is
\[ \func{QFI}{\ket{SMSV(\phi,t)}_V} = 2\left(t \omega \func{sinh}{2r}\right)^2\]
which can be greater than the single photon state when $r>1/2$. As per the introduction, the reason for this increased quantum Fisher information is because this state is a superposition over states with higher energy than the single photon state.\\

I have several responses to this observation.

First. Why doesn't LIGO ever talk about this state as containing higher energy components with shorter wavelengths? Why doesn't LIGO ever present evidence of this state as having higher frequency and shorter wavelength components? That is precisely the basis behind the higher quantum Fisher information of this state.

Second, $\ket{SMSV}$ is a pure state that spans a Hilbert space with unbounded energy support. And it is a pure state with a well-defined phase relationship $\theta$ in the complex amplitude of each basis state. The creation of such a state requires the creation of a Hilbert space with infinite energy. Even if this state is truncated at some finite $n = n_0$, the creation of this state with well-defined complex amplitudes for every basis vector requires generation of an $n_0$ photon state (more generally a Hilbert space with the same number of basis vectors and same energy). The evidence for creation of such a state requires measurements of this phase relationship between basis states. Simply presenting photon counting statistics is not evidence of this state since they can be reproduced with a mixed state that is a probability distribution over different number states.

Third. If this state was being generated, then it can be inserted directly into just one of the arms of LIGO and it would be sensitive to a change in length in that arm. No beam-splitter is needed to detect a relative path length difference. I have never seen any discussion mentioning that this is possible.

Fourth. If this state was used to detect a phase difference, this would be evidenced by observing a greater change in the measurement outcomes for the same phase difference. The key property of this state, which is not present in any data shown by LIGO, is an $n$-fold increased phase shift experienced by the $\ket{2n}$ basis states. This is the key property of any quantum state with higher quantum Fisher information. Observation of less measurement noise is not a hallmark of this state.\\

As a general rule, measurement precision can be calibrated either by comparing how close the estimate is from the true value of a known parameter, or by providing evidence of how the measurement results change in response to the signal. LIGO never presents this evidence, instead they focus only on the noise. This is such an obvious oversight that I am stunned it has gone overlooked for so long.

\subsection{Why do you restrict your analysis to (Fisher) information and not the actual measurement uncertainty? Are you hiding something behind this definition?}

I analyse the information and not the measurement uncertainty for a reason that is both somewhat uninteresting and at the same time points to a bigger issue in quantum metrology. The reason is that it is easy to violate bounds on the measurement uncertainty, whereas the same is not true for the information. This makes the uncertainty bound non-rigorous.

In general, the following inequality is used to relate the information (from a single measurement) to the estimation uncertainty \cite{Braunstein1996, Giovannetti2004, Giovannetti2006, DemkowiczDobrzanski2012, Pang2014, Yuan2015, Pang2017}:
\be
\langle (\Delta \est{\theta})^2\rangle \ge \frac{1}{\func{I}{\theta,t}},
\label{eq:info}
\ee
where $\langle (\Delta \est{\theta})^2\rangle$ is the expected mean squared error of the estimator, i.e. the uncertainty $\Delta \est{\theta}$ is the positive square-root of the expected mean squared error of the estimator.

For $t = 0$, we have that $\func{I}{\theta,t} = 0$. No information is obtained on the signal $\theta$ and the uncertainty is infinite. However any prior information on $\theta$, even just defining the allowable values that $\theta$ can take, means the initial uncertainty is finite and necessarily violates this bound. To account for prior information we could reformulate our uncertainty definition and work in terms of relative uncertainty reduction. Now we could say that Eq.\,\eqref{eq:info} relates the amount of information provided by a measurement to the \textit{relative} uncertainty reduction (with respect to the prior uncertainty). However, even using this definition of relative uncertainty reduction it is possible to violate the bound. Take the following discrete prior probability distribution of $\theta$:
\[ \func{Pr}{\theta = \pi/2} = 1/2; \quad \quad \func{Pr}{\theta = \pi} = 1/2, \]
i.e. $\theta$ can take one of two values with equal probability. We have an initial uncertainty in $\theta$ of $\pi/4$, and we want to calculate an uncertainty reduction per unit time relative to this initial value. For this prior, the bound on the relative uncertainty reduction given by Eq.\,\eqref{eq:info} can be violated by an arbitrary amount. To see this, we can perform a measurement where a single spin sensor is driven to one of two orthogonal states (e.g. $\ket{0}$, $\ket{1}$) dependent on the value of $\theta$; so that if $\theta = \pi/2$, the spin ends up in $\ket{0}$ and if $\theta = \pi$, the spin ends up in $\ket{1}$ (see \cite{Helstrom1976, Childs2000, Aharonov2002} for detailed discussion). If readout of the spin state is perfect, then regardless of how long it takes to perform unitary evolution, the final uncertainty on $\theta$ is zero. Even allowing for non-perfect readout, as long as the probability to obtain a correct result is $>1/2$, the measurement can be repeated, thus producing an exponential reduction in uncertainty. We can prove these methods violate the bound in Eq.\,\eqref{eq:info} because the uncertainty is at least an exponentially reducing function in time (for arbitrarily long times), whereas $\func{I}{\theta,t}$ is at most a quadratically increasing function of $t$.

These examples show that to obtain a rigorous bound on the measurement uncertainty, we need to be careful in considering the prior information, because the final uncertainty depends on both the information provided from the measurement and our prior uncertainty. Performing the above analysis adds complexity, is distracting and forces us to put restrictions on the prior probability distribution to avoid pathological priors. This makes the analysis less general. Most importantly the above analysis does not address the issue at stake. We don't care about the prior probability distribution and how that impacts the final measurement uncertainty, what we want to know is the following:
\begin{quotation}
Do measurements of entangled squeezed states give more information on a signal than measurements of unentangled states?
\end{quotation}
Perhaps when converting from information to uncertainty, the result may violate the bound of Eq.\,\eqref{eq:info}. But what we can say with confidence, is that: if the answer is no, then measurements of entangled squeezed states will not provide a lower uncertainty than measurements unentangled states (see also the discussion in Appendix.\,\ref{App:Assumptions}).\footnote{An earlier version included this explanation which I no longer agree with ``Mathematically, the reason why we can violate Eq.\,\eqref{eq:info} -- known as the Cram\'er-Rao bound -- in the above example is because the posterior probability distribution does not satisfy the regularity condition 
\[ \8{\frac{\partial \mathrm{ln}\left(\func{Pr}{x;\theta,t}\right)}{\partial \theta}} = 0 \quad \mathrm{for \; all}\;\theta\]
where the expectation is taken with respect to $\func{Pr}{x;\theta,t}$ (see section on Cram\'er-Rao lower bound in \cite{Kay1993}). This condition is assumed in the derivation of the uncertainty bound on the mean squared error of the estimator. If we restrict ourselves to prior probability distributions that ensure the posterior probability distribution satisfies the regularity condition, then we can prove that the bound holds."}

A final point, similar analysis has been used to claim violations of the Heisenberg time-energy uncertainty relation \cite{Aharonov2002} and used as evidence that quantum computers can exponentially outperform classical computers \cite{Atia2017}. To me the above arguments show that this should actually be interpreted as evidence that quantum computers cannot outperform classical computers.\\

\subsection{What about the Heisenberg uncertainty relations for conjugate observables. Don't they disprove this work. Are you saying they are wrong?}\label{FAQ:3}

Instead of operating in the Schr\"odinger picture we could analyse how unitary evolution acts as a mapping of quantum states in the Heisenberg picture. Then, for $\Op{H}(\theta, t) = \theta \Op{H}$, we have:
\[\frac{\mathrm{d} \Op{U}(\theta,t)}{\mathrm{d}\theta}\ket{\psi_0} = -\iu t \Op{H} \, \Op{U}(\theta, t)/\hbar \ket{\psi_0}.\]
As $\Op{U}$ and $\Op{H}$ commute, expressing $\ket{\psi_0}$ in terms of the eigenstates of $\Op{H}$, we have:
%\begin*{align}
\[
\begin{split}
& \left(\frac{\mathrm{d}\bra{\psi(\theta,t)}}{\mathrm{d} \theta}\right)\left(\frac{\mathrm{d}\ket{\psi(\theta,t)}}{\mathrm{d} \theta}\right) = \bra{\psi_0}\left(-\iu t \Op{H} \, \Op{U}(\theta,t)/\hbar\right)^{\dagger}\left(-\iu t \Op{H} \, \Op{U}(\theta,t)/\hbar\right)\ket{\psi_0} \\
& = (t/\hbar)^2 \sum_{k = 1}^K \bra{\psi_{E_k}} \left(\func{Exp}{\iu \theta t E_k/\hbar}E_k \alpha_k^{\ast}\right)\left(\alpha_k E_k\func{Exp}{-\iu \theta t E_k/\hbar}\right) \ket{\psi_{E_k}}\\
& = (t/\hbar)^2 \sum_{k = 1}^K  E_k^2 |\alpha_k|^2,\\
\end{split}
\]
and
\[ \left|\bra{\psi(\theta,t)}\left(\frac{\mathrm{d}\ket{\psi(\theta,t)}}{\mathrm{d} \theta}\right)\right|^2 = \left|-\iu t/\hbar \bra{\psi_0}\Op{U}^{\dagger} \Op{H} \,\Op{U}\ket{\psi_0}\right|^2 = (t/\hbar)^2\left(\sum_{k = 1}^K  E_k |\alpha_k|^2\right)^2.\]
%\end*{align}
Thus, for this Hamiltonian, maximising information on $\theta$ means maximising the variance of $\Op{H}$: $\func{Var}{\Op{H}} \equiv \langle \Op{H}^2 \rangle - \langle \Op{H} \rangle^2$ with respect to $\ket{\psi_0}$. For unitary evolution this also means maximising the variance of $\Op{H}$ on the output state.\footnote{In general we want to maximise the variance of the generator of parameter translation (see \cite{Braunstein1996, Pang2014, Pang2017}). For a time-independent Hamiltonian, if $\Op{U}$ and $\partial \Op{H}(\theta)/\partial \theta$ commute, then we want to maximise the variance of $\partial \Op{H}(\theta)/\partial \theta$.} We have:
\[ \func{I}{\theta} = \frac{4t^2}{\hbar^2}\func{Var}{\Op{H}}.\]

Due to the correspondence between different pictures, increasing $\func{Var}{\Op{H}}$ in this context only comes about through increasing the state response to $\theta$. We can also see that parameter estimation is equivalent to energy estimation for this Hamiltonian (with an appropriate units conversion factor). %\textcolor{blue}{In this analysis, it is incorrect to equate $\sqrt{\func{Var}{\Op{H}}}$ as the uncertainty in estimating the energy eigenvalues or the Hamiltonian $\Op{H}$. It is the variance of the measurement results, assuming perfect measurements.}
Importantly, we see that states with large variance in measurement outcomes (i.e. noise) give the most information on $\theta$. This is in direct contradiction with the story promoted by the squeezing community.

There are some technical aspects to this analysis which means that working in the Schr\"odinger picture is preferable. In the main text, we assume that two state-vectors have the same dependence on $\theta$. However in the Heisenberg picture, the dependence is included in the operator not the state-vector, so we cannot make that comparison. More pertinently, it is possible to mistakenly assume that the statistics of real measurements must perfectly reflect the properties of the mathematical object -- the measurement operator -- associated with that measurement. In particular, the experimentalist who asserts they perform a measurement corresponding to a mathematical operator, always has the freedom to ``cheat", by altering the measurement statistics via post-processing to either reduce or increase the measurement variance as they like.

This freedom is implied because processing of raw experimental data is inherent in quantum measurements and is required to obtain estimators. Experimentally one does not measure energy by mathematically applying the $\Op{H}$ operator to a system upon which the outcome $E_k$ is returned. Instead, particularly in spin experiments, spins are readout fluorescently where photons are collected and differences in photon counts are related to measurement outcomes. There is nothing to stop one from processing the photon data and artificially increasing the variance, or for that matter either not collecting data or performing disingenuous subtractions to reduce the variance. These operations do not improve the measurement precision and it can be hard to identify such artificial enhancements and separate them from actual improvements. If we simply accept claims based on the measurement variance, then it can be difficult to disentangle how the measurement statistics reflect the information on the parameter.

Take the following phase estimation experiments. In one experiment, the state to be measured is $\left(\ket{1} + e^{\iu \theta}\ket{0}\right)/\sqrt{2}$. Compare to an experiment with the entangled Bell state $\left(\ket{1 1}+ e^{\iu \theta}\ket{0 0}\right)/\sqrt{2}$. What are the uncertainty bounds on estimating $\theta$ from these two states? The measurement outcomes of the second experiment produce twice as many photons as the first and the variance of the Bell state with respect to an $\Op{S}_z$ operator is greater than the single particle state. But in the limit of perfect fidelity readout, these extra photons do not provide more information on $\theta$ -- they are correlated! Squeezing people might actually claim that because the $\Op{S}_z$ observable from the Bell state has twice the variance of the single spin state, then it has a worse uncertainty whereas by measuring a conjugate observable a better uncertainty can be obtained. That overlooks the fact that the two states are isomorphic, assuming the postulates of quantum mechanics, both states provide the same amount of information on $\theta$ (see the comment on Assumption 1 in Appendix\,\ref{App:Assumptions}). The variance argument presented in spin-squeezing is incorrect in many respects, for one it is a simple rescaling error. But it is difficult to catch. The analytic mistake introduced by looking at variances of observables becomes even harder to identify when we start talking about measurements on ensembles where the contribution of each state gets blurred out. It is much easier to pick up on this error when comparing individual normalised states in the Schr\"odinger picture.

A final point. It is possible to rigorously bound the uncertainty on estimating a parameter $\theta$ from the measurement variance of some datum $x$. To do this, two further pieces of information are required: i) Quantification the data dependence on the parameter, $\partial \func{Pr}{x;\theta}/\partial \theta$ where the gradient is taken at the actual value of $\theta$. ii) Quantification of the estimator bias. After reading through the squeezing papers referenced in the main text, I would say that in general, squeezing literature is extremely misleading when it comes to discussing these two points. In my opinion, the lesson of the squeezing literature is not a physics lesson, it is not about how ``entanglement allows quantum states with less variance to be produced", rather it is a cautionary lesson on how easy it is to be misled based on incorrect statistical analysis.

\end{document}